\documentclass[twocolumn]{aastex6} 
\usepackage{graphicx}
\usepackage{epstopdf}
\usepackage{natbib}
\usepackage{rotating}
\usepackage{makecell}
\usepackage{morefloats} 
\def\ie{{\it i.e., }}
\def\bea{\begin{eqnarray}}
\def\eea{\end{eqnarray}}
\newcommand{\eg}{{\it e.g., }} 
\newcommand{\HI}{\hbox{{\rm H}\kern 0.1em{\sc i}}}
\newcommand{\HII}{\hbox{{\rm H}\kern 0.1em{\sc ii}}}
\newcommand{\NII}{\hbox{{\rm N}\kern 0.1em{\sc ii}}}

\begin{document}

\title{CHANG-ES XVII: H$\alpha$ Imaging of Nearby Edge-on Galaxies, New SFRs, and an Extreme Star Formation Region -- Data Release 2 }

\author{Carlos J. Vargas\altaffilmark{1,2}, Ren\'{e} A. M.  Walterbos\altaffilmark{2}, Richard J. Rand\altaffilmark{3}, Jeroen Stil\altaffilmark{4}, Marita Krause\altaffilmark{5}, Jiang-Tao Li\altaffilmark{6}, Judith Irwin\altaffilmark{7}, Ralf-J\"{u}rgen Dettmar\altaffilmark{8}}

\altaffiltext{1}{Department of Astronomy and Steward Observatory, University of Arizona, Tucson, AZ, U.S.A.}
\altaffiltext{2}{Department of Astronomy, New Mexico State University, Las Cruces, NM 88001, U.S.A.}
\altaffiltext{3}{Department of Physics and Astronomy, University of New Mexico, 1919 Lomas Blvd. NE, Albuquerque, NM 87131, U.S.A.}
\altaffiltext{4}{Department of Physics and Astronomy, University of Calgary, Canada}
\altaffiltext{5}{Max-Planck-Institut f\"{u}r Radioastronomie, Auf dem H\"{u}gel 69, 53121 Bonn, Germany}
\altaffiltext{6}{Department of Astronomy, University of Michigan, 311 West Hall, 1085 S. University Ave, Ann Arbor, MI, 48109-1107, U.S.A}
\altaffiltext{7}{Department of Physics, Engineering Physics, \& Astronomy, Queen’s University, Kingston, Ontario, Canada, K7L 3N6}
\altaffiltext{8}{Ruhr-University Bochum, Faculty of Physics and Astronomy, Astronomical Institute, 44780 Bochum, Germany}

\begin{abstract}

We present new narrow-band H$\alpha$ imaging for $24$ nearby edge-on galaxies in the CHANG-ES survey. We use the images in conjunction with WISE $22$ micron imaging of the sample to estimate improved star formation rates (SFRs) using the updated recipe from \citet{vargasetal18}. We explore correlations between the updated star formation properties and radio continuum scale heights, scale lengths, and diameters, measured in \citet{krauseetal18}. We find a newly discovered correlation between SFR and radio scale height that did not exist using mid-IR only SFR calibrations. This implies that a mid-IR extinction correction should be applied to SFR calibrations when used in edge-on galaxies, due to attenuation by dust. The updated SFR values also show newly discovered correlations with radio scale length and radio diameter, implying that the previously-measured relationship between radio scale height and radio diameter originates from star formation within the disk. We also identify a region of star formation located at extreme distance from the disk of NGC 4157, possibly ionized by a single O5.5 V star. This region is spatially coincident with an XUV disk feature, as traced by GALEX NUV imaging. We theorize that the star formation feature arose due to gravitational instability within gas from an accretion event. New H$\alpha$ images from this work can be found at the CHANG-ES data release web site, https://www.queensu.ca/changes.

\end{abstract}

\section{Introduction}

Observations of edge-on galaxies are ideal for studying interactions between matter within a spiral galaxy's star-forming disk and extraplanar matter outside of the disk. Edge-on observations of the extraplanar diffuse ionized gas allow study of the global influence of released energy from young, massive stars into and out of the interstellar medium (ISM), near the disk-halo interface.  

NGC 891 was the first galaxy, other than the Milky Way, observed to harbor an extraplanar component of diffuse ionized gas (DIG; \citealt{dettmar90,randetal90}). Observational evidence of extraplanar DIG in numerous other galaxies soon followed (e.g. \citealt{pildisetal94,rand96,hoopesetal99,rossadettmar00,milleretal03}). 

Active star formation may lead to hot outflows from supernovae and stellar winds away from disks and into the halo. In a more classical picture this is described in terms of chimneys or superbubbles (see, {\eg}\citealt{normanikeuchi89}). This scenario works well in the case of localized, very high star formation rates as seen in galaxies with nuclear starbursts where this mechanism is supported observationally ({\eg} \citealt{ceciletal02}; \citealt{stricklandetal04}) as well as theoretically \citep{maclowetal99}.

There is less observational evidence for such ``outflow cones" in `normal' star-forming galaxies. For example, \citet{normanikeuchi89} predict a much larger occurrence of chimneys/superbubbles than was discovered in ground-based \citep{howksavage00} or space-based \citep{rossaetal04} studies of the DIG distribution in edge-on galaxies. This is most likely due to the fact that the relevant time scales for star formation and heating and cooling lead to a more dynamical and thus highly structured ISM as described by numerical simulations \citep{kimostriker18,hilletal18,gentetal13}. The models predict hot gas outflows as seen in `normal' star-forming galaxies through the detection of diffuse interstellar X-ray emission above the disk ({\eg} \citealt{wangetal01}) and they can also explain the association of {\HI}  holes in galaxy disks with high velocity gas \citep{boomsmaetal08}. The structuring of the ISM into sheets, bubbles, and holes create pathways for ionizing photons to escape the dense {\HI} disk environment. 

Hot, massive stars are the primary source of DIG ionization, but cannot explain all the line ratio behavior, and other contributing processes have been examined (see review by \citealt{haffneretal09}). Hot, evolved low mass stars have recently received some attention as a secondary contributor. For example, a study by \citet{randetal11}, which included crucial information from mid-infrared line ratios, explored whether the DIG in NGC 891 may arise from a thin disk of massive stars in combination with a thick ($\sim1$ kpc scale height) disk of hot evolved stars. However, the inclusion of this second component does not completely account for the deficiencies of massive star ionization models in predicting line ratios. Also, uncertainties in the number, temperatures, and distribution of this component make it very difficult to characterize its contribution to DIG ionization and heating. However, the presence of DIG with low H$\alpha$ equivalent widths in bulges and early type galaxies (e.g. \citealt{lacerdaetal18}; \citealt{stasinskaetal08}) indicates that hot, evolved stars should indeed be a contributing source.

The Continuum Halos in Nearby Galaxies -- an EVLA Survey (CHANG-ES; \citealt{irwinetal12, irwinetal12b, irwinetal13, wiegertetal15}) aims to establish the connection between star formation in galaxy disks and non-thermal processes, such as halo magnetic fields, cosmic ray (CR) injection and propagation, and active galactic nuclei (AGN). The CHANG-ES survey involves Karl G. Jansky Very Large Array (VLA) observations in B, C, and D configurations of 35 highly-inclined nearby galaxies. The observations encompass two VLA bands; L-band ($1-2$ GHz) and C-band ($5-7$ GHz) in all polarization products. The CHANG-ES VLA radio continuum data are, on average, an order of magnitude improvement in sensitivity to previous observations. 

Radio continuum emission from star forming galaxies contains both free-free (thermal) and Synchrotron (non-thermal) emission. A method of separating each component from the ensemble is needed to study each individually. Where data from only two discrete radio continuum bands are present, separation of the two components has traditionally been carried out by assuming a constant non-thermal spectral index (e.g. \citealt{kleinetal82}). Since this method makes studying variations in the non-thermal spectrum impossible, further separation methods have been devised. 

\citet{becketal82} were among the first to estimate the thermal (free-free) emission from M31 using a catalog of {\HII} regions, corrected for extinction. Since both free-free emission and recombination line emission originate from ionized regions, recombination line emission is a good tracer of thermal radio emission. H$\alpha$ emission, in particular, is easily observed from nearby galaxies at optical wavelengths, and is the strongest Balmer line. However, H$\alpha$ emission suffers from strong attenuation by foreground dust, and thus requires a correction for this extinction. This problem is exacerbated in the edge-on perspective. We note that radio recombination lines would be ideal for inferring thermal emission, since they would not suffer from extinction. These lines are unfortunately too faint to map on a resolved basis with existing technology. \citet{tabatabaeietal07} attempted to map radio recombination lines in M33 with Effelsberg $6.2$ cm observations, but were unsuccessful. Carbon recombination lines have been detected with LOFAR in \citet{morabitoetal14}, but were associated with cold atomic gas near the nucleus of M82.

Existing empirical relations between mid-IR emission from heated dust grains and star formation link mid-IR emission to thermal radio emission. This mid-IR emission can be further related to the extinction of H$\alpha$ emission and used as a SFR tracer. Hence, star formation in both obscured and un-obscured regions can be estimated by a combination of H$\alpha$ and $24$ micron measurements \citep{kennicuttetal07}. The robustness of this relationship has been established using Pa$\alpha$ emission \citep{calzettietal07}, and was further tested with integrated measurements of galaxies \cite{kennicuttetal09}.    

Modern studies of independent thermal radio continuum component estimation exist in spiral galaxies at various inclinations \citep{tabatabaeietal07,basuetal12,leroyetal12,basuetal17}. However, the edge-on perspective adds complexity to this problem. Line-of-sight extinction is certainly largest in the mid-planes of edge-on galaxies due to the increased path length of photons traveling through the disk. Thus, the observed H$\alpha$ emission may be only from sources on the near-side of the galaxy. Additional non-homogeneities, like spiral arms, bars, and distortions make observed H$\alpha$ emission in edge-on galaxies not necessarily representative of the full disk emission. This necessitates the inclusion of mid-IR emission of edge-on galaxy mid-planes, which is less subject to, but not completely unaffected by extinction. 

A method of predicting the thermal radio component using a mixture of H$\alpha$ emission and mid-IR emission specifically for edge-on galaxies was introduced in \citet{vargasetal18}. This method builds upon empirical star formation rate calibrations \citep{calzettietal07,jarrettetal13} by taking into account mid-IR extinction that would otherwise lead to underestimates within the edge-on disk. Star formation rate (SFR) calculations using this method are not affected by the presence of AGN, which is the case for radio-based estimates. The method also relies heavily on the inclusion of H$\alpha$ imaging, which is particularly important in estimating the predicted thermal radio emission outside of the galaxy disks. Thus, consistent H$\alpha$ observations of the CHANG-ES sample is needed for thermal prediction, and to estimate disk star formation properties. Such observations are presented here.

L-band VLA Radio continuum emission from a sample of six edge-on galaxies in the Sydney AAO Multi-object Integral Field Spectrograph (SAMI) Galaxy Survey \citep{croometal12,bryantetal15} were studied closely in \citet{leslieetal17}. That study found that extended radio continuum halos correlate with optical galactic wind-like features, as identified through velocity field and dispersion critera outlined in \citet{hoetal16}. The galaxies in their sample with extended radio continuum halos also show high specific SFRs and high SFR surface densities, implying that star formation, CR emission, and outflows are all connected. Recent numerical simulations conclude that the observed global properties of the ISM in disk galaxies such as the scale height perpendicular to the disk can only be explained if the effects of CRs are taken into account. It is suggested that CRs are  a significant contribution to launching galactic winds \citep{hanaszetal13,boothetal13,salemetal14}. The H$\alpha$ imaging presented in this study, in conjunction with the CHANG-ES radio continuum data, provide an opportunity to study the relationship between star formation and CR injection in more detail. 

This study expands on previous H$\alpha$ imaging of edge-on galaxies in the literature, and establishes a new set of H$\alpha$ images. Only $\sim10$ CHANG-ES galaxies have adequate H$\alpha$ imaging of their entire disks in the literature. A complete set of H$\alpha$ images matching the CHANG-ES sample is needed to study the DIG distribution and estimate SFR. Furthermore, a future study is planned utilizing this complete set of images to independently estimate the thermal radio component in all CHANG-ES galaxies. We use the SFRs from the H$\alpha$ imaging of the full CHANG-ES sample to further explore the validity of the $22$ micron extinction correction introduced in \citet{vargasetal18}, and to explore the relationship between SFR and the scale height of radio continuum emission.

\section{Observations and Data} 

We used the Apache Point Observatory (APO) 3.5-m Telescope and the Astrophysical Research Consortium Telescope Imaging Camera (ARCTIC) to obtain new observations of 25 of the 35 CHANG-ES sample galaxies. The unbinned plate scale of ARCTIC is $0.114 \arcsec$ per pixel, and we used a $2\times2$ pixel binning, giving the resulting images $0.228\arcsec$ pixels. The field of view of the instrument spans $7.5 \arcmin$ square. The large field of view is reasonably matched to the nearby galaxies in the CHANG-ES sample. The $10$ remaining CHANG-ES galaxies have pre-existing H$\alpha$ imaging that is usable for the purposes of this study. The observations spanned December 2015 through December 2016. For each field three exposures lasting $3$ minutes each were taken in a broad-band SDSS r-band filter, and three exposures lasting $15$ minutes each were taken in an optimally-redshifted narrow-band (NB) H$\alpha$ filter. The FWHM of the narrow-band filter is $30\rm{\AA}$. We observed nearby standard stars from \citet{oke90} for flux calibration in photometric weather.  Weather, humidity, and time considerations made dome flats a necessity over twilight sky flats on some nights. Also, some observations were taken in non-photometric weather, as determined by sky transparency variations measured by the APO infrared sky camera at the time of observations. The field of view is slightly vignetted due to low levels of light scatter out of the field of view near the edges. The sensitivity of each resulting continuum-subtracted H$\alpha$ image is included in Table \ref{obstable}, alongside other observational parameters. Surface brightnesses were converted to emission measure (EM) assuming gas at 10,000 K, via 1 pc cm$^{-6}$ $=$ $2\times 10^{-18}$ erg cm$^{-2}$ s$^{-1}$ arcsec$^{-2}$.

\begin{table*}[h]
\addcontentsline{lot}{table}{ARCTIC H$\alpha$ Observation Details for CHANG-ES Galaxies}
\tiny
\centering

    \begin{tabular}{c c c c c c c c}
    \hline
    \hline
    Galaxy & Date Observed & Number of fields & NB Filter & Flat Type & Photometric & $f_{{\NII}}$ & EM Noise (pc cm$^{-6}$)\\ 
    \hline
    NGC 660 &  Oct. 1, 2016 & 2 & 658/3 & Dome & No & 0.15 & 2.95 \\ 
    NGC 891 &  N/A \\
    NGC 2613 &  Dec. 27, 2016 & 1 & 660/3 & Dome & Yes & 0.30 & 3.76 \\
    NGC 2683 &  Dec. 10, 2015 & 2 & 658/3 & Twilight & No & 0.10 & 6.13 \\
    NGC 2820 &  Dec. 27, 2016 & 1 & 660/3 & Twilight & Yes & 0.25 & 5.25\\
    NGC 2992 &  Apr. 3, 2016 & 1 & 661/3 & Twilight & Yes & 0.20 & 3.53\\
    NGC 3003 &  Dec. 10, 2015 & 1 & 659/3 & Twilight & No & 0.10 & 3.21\\
    NGC 3044 &  Dec. 10, 2015 & 1 & 659/3 & Twilight & No & 0.15 & 10.2 \\
    NGC 3079 &  Dec. 27, 2016 & 2 & 659/3 & Dome & Yes & 0.10 & 4.33\\
    NGC 3432 &  Jan. 12 \& June 6, 2016 & 2 & 658/3 & Dome & No & 0.20 & 2.72 \\
    NGC 3448 &  Apr. 3, 2016 & 1 & 659/3 & Twilight & Yes & 0.20 & 4.70\\
    NGC 3556 &  Jan. 12, 2016 & 2 & 658/3 & Twilight & No & 0.20 & 2.88\\
    NGC 3628 &  May 27, 2016 & 3 & 658/3 & Both &  No & 0.35 & 11.9 \\ 
    NGC 3735 &  May 27, 2016 & 1 & 661$^\star$ & Dome & Yes & 0.5 & 4.07\\
    NGC 3877 &  Feb. 13, 2016 & 1 & 658/3 & Twilight & No & 0.25 & 6.37\\
    NGC 4013 & Jan. 12, 2016 & 1 & 658/3 & Twilight & No & 0.25 & 6.77\\
    NGC 4096 & Feb. 13, 2016 & 2 & 658/3 & Twilight & No & 0.20 & 6.43\\
    NGC 4157 & Feb. 13, 2016 & 2 & 658/3 & Twilight & No & 0.25 & 37.1 \\
    NGC 4192 & Mar. 13, 2016 & 2 & 657/3 & Dome & Yes & 0.15 & 1.03\\
    NGC 4217 & N/A \\
    NGC 4244 & N/A \\
    NGC 4302 & N/A \\
    NGC 4388 & Apr. 3, 2016 & 1 & 661/3 & Twilight & Yes & 0.30 & 3.41\\
    NGC 4438 & Mar. 13, 2016 & 2 & 657/3 & Dome & Yes & 0.15 & N/A\\
    NGC 4565 & N/A \\
    NGC 4594 & N/A \\
    NGC 4631 & N/A \\
    NGC 4666 & Dec. 27, 2016 & 1 & 660/3 & Dome & Yes & 0.10 & 3.33\\
    NGC 4845 & Apr. 9, 2016 & 1 & 659/3 & Dome & Yes & 0.20 & 4.59\\ 
    NGC 5084 & N/A \\
    NGC 5297 & Apr. 9, 2016 & 1 & 661/3 & Dome & Yes & 0.20 & 4.07\\
    NGC 5775 & N/A &  & \\
    NGC 5792 & May 27, 2016 \& Jul. 6, 2016 & 2 & 661/3 & Dome & No & 0.15 & 4.48 \\
    NGC 5907 & N/A \\
    UGC 10288 & Jul. 6, 2016 & 1 & 661/3 & Dome & No & 0.15 & 6.35\\

    \hline   
    \end{tabular}
    
\caption{ARCTIC H$\alpha$ observation details of CHANG-ES galaxies. Galaxies with `N/A' listed for the observation date were not observed, but have existing H$\alpha$ imaging in the literature (exception: NGC 5084). Galaxies observed in photometric conditions were flux calibrated using standard stars, and galaxies not observed in photometric conditions were calibrated using SDSS photometry of stars within the field. The column, $f_{{\NII}}$ is an estimate of the fraction of the total {\NII} line strength that is within the used narrow-band filter.\\
$\star$ A smaller 2-inch narrow-band filter was used for this galaxy. }
\label{obstable}
\end{table*}

\subsection{Image Reductions}

\begin{figure*}
\centering
\includegraphics[scale=0.47]{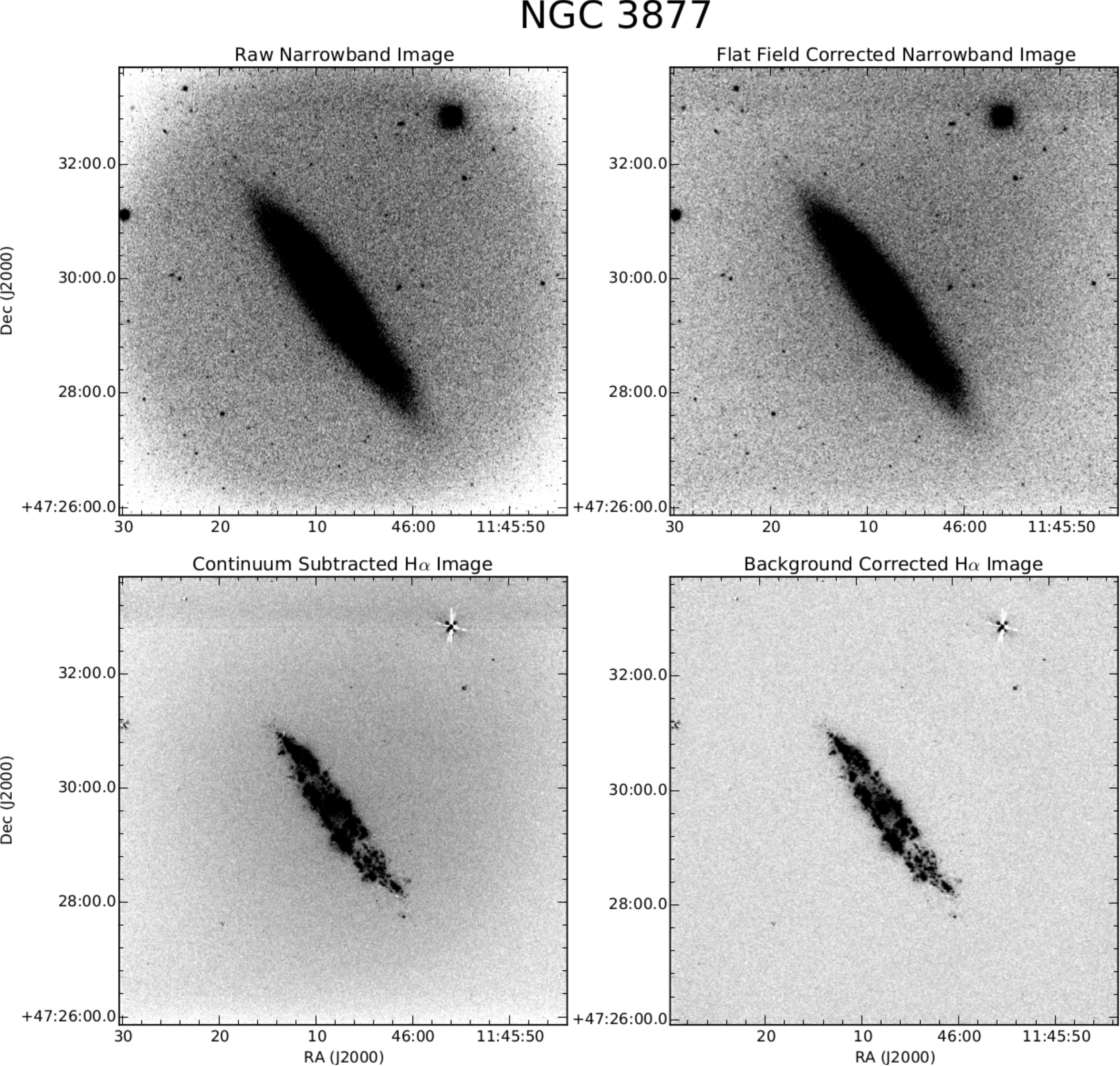}
\caption[Data Reduction Steps]{Data reduction steps concerning background variations for NGC 3877. Upper left: raw narrow-band image. Upper right: narrow-band image after applying the flat field correction. Lower left: continuum subctracted H$\alpha$ image. Lower right: continuum subtracted H$\alpha$ image after applying the background fitting correction. All images are shown in a logarithmic stretch.}
\label{datareductionplot}
\end{figure*}

The APO imaging was reduced using the Image Reduction and Analysis Facility (IRAF; \citealt{tody86}) and packages therein. Bias overscan regions were fit, corrected for, and trimmed using \textit{CCDPROC}. A combined bias image was used to bias correct each image, and was created from at least $22$ individual bias images. The combined bias frame was used to correct each image using \textit{CCDPROC}. Flat fields were taken in each night; of the twilight sky, or dome with dim quartz lamps (see Table \ref{obstable}). These flats were then combined and used to correct each image using \textit{CCDPROC}. Figure \ref{datareductionplot} follows the major steps taken in the image reductions for NGC 3877, as an example. 

Since these observations were all taken early in the commissioning of the ARCTIC instrument, some problems arose in the reduction stages. The bias overscan regions did not show uniform characteristics frame-to-frame. In flat field images, the bias overscan seemed to increase non-linearly at low row number, while the overscan in the bias frames was linear across all rows. Furthermore, the bias overscan in galaxy frames typically showed sharp jumps at randomly-occuring rows. The overscan regions in most frames were not adequately fit with a polynomial of order 2. So, a polynomial of order 3 was used to fit the bias overscan regions. The higher order fit is an improvement, but does not perfectly remove the erratic jumps in the bias overscan. Furthermore, flat fielding of the narrow-band images is less effective than for the r-band images. This likely arises from excess scattered light having a more pronounced effect in the narrow-band interference filter, than on the broad-band r filter. To correct these issues, we use \textit{BACKGROUND} to fit an order 5 Legendre polynomial to the background of each final continuum-subtracted H$\alpha$ image, along either rows or columns, depending on the galaxy geometry. This background fit is then subtracted from the image to correct for the excess scattered light. Care was taken to exclude emission from the galaxy during the fitting process by rejecting pixel values $>2\sigma$ from the fit. 

We analyzed the effects of the background fitting routine on the total flux of a given galaxy by measuring the total flux both before and after the background fitting routine is applied. The excess background near the galaxy was estimated and subtracted in the images without the background fitting routine applied. These total flux measurements vary by only $\sim1\%$ in most cases, and up to $\sim3\%$ in one case (NGC 4666). Hence, uncertainties in the background fitting procedure can only affect the H$\alpha$ intensity and SFR estimates by $\sim1\%$.     

After bias corrections and flat field corrections, images were shifted to align using the centroid positions of numerous unsaturated stars in both the narrow-band and continuum filter images. A linear interpolation was used to shift images accurate at the sub-pixel level. After shifting, each individual image's resolution was matched by Gaussian smoothing to the image with the largest PSF. Cosmic ray events were removed from individual images using the code `Detect and Remove Cosmic Rays' (DCR; \citealt{pych04}). This code calculates image statistics in a definable-sized box which moves to sample the entire image and rejects pixels above a threshold in intensity. The code can sometimes clip the edges of stars, so great care was taken to define cosmic ray detection parameters that avoid this. This conservative approach led to some remaining cosmic rays in each image. To remove these, the images were combined with \textit{IMCOMBINE}, utilizing `minmax' rejection. Only images with equal exposure time were combined. Care was also taken to equalize the sky level for each image prior to combination. 

The narrow-band images were continuum subtracted using the r-band continuum filter images. The combined r-band image was scaled to match the intensity of the narrow-band image by a factor determined by photometry of stars in each image. After scaling, the r-band images were subtracted from the narrow-band image, leaving only H$\alpha$ line emission. The r-band scaling factor was iterated on after visual inspection of the resulting continuum-subtracted H$\alpha$ image to mitigate obvious over-subtraction features, such as dark negative regions around the edges of galaxy emission.  The r-band images for NGC 4438 all needed to be discarded due to excess twilight light, so that galaxy is not currently included in the analysis. Due to time constraints because of poor weather, NGC 5084 was not observed. NGC 5084 has a low SFR as inferred by IRAS total infrared observations, and faint radio continuum emission, so we do not expect to find a considerable amount of strong H$\alpha$ emission in that galaxy. 

As seen in Figure \ref{datareductionplot}, the vignetting in the field is clearly seen in the raw image as darkness at the edge of the field. Imperfections in the background remain near the edges of the field after the flat-field correction, which persist in the continuum subtracted H$\alpha$ image. The background fitting removes variations in the background, and sets the background level to have a mean of zero.  

A world coordinate system (WCS) was assigned to each image using the following method. The physical positions of stars in each image were calculated using a centroid fit to each star, and compared to the quoted physical coordinates of those same stars in an existing SDSS r-band image. The positions of at least $10$ stars (often more) were used as input to \textit{CCMAP} (within IRAF), which creates a plate solution using the physical and image pixel positions of stars. The WCS solution is then applied to each image using \textit{CCSETWCS}, also within IRAF.   

In galaxies with two or more observational fields, each field's continuum-subtracted H$\alpha$ images were combined using Montage \citep{berrimanetal03}. The combination was used with `background matching'. However, obtaining a mosaic image with well-matched backgrounds is challenging due to the aforementioned vignetting and flat fielding issues with the narrow-band filter. To quell this issue, each field is trimmed to exclude the vignetted regions, and only images that have been background fit are used for mosaicing. Despite the trimming, most fields had significant overlapping regions, within which the average pixel value was used for the mosaic image.  

The flux is calibrated in each narrow-band image using spectrophotometric standard stars from \citet{oke90} that were observed on the same night as the galaxy observations, when the weather was photometric. When the weather was not photometric, the flux was calibrated using quoted SDSS r-band magnitudes of stars in the field. At least $3$ stars were used in each SDSS r-band field. The flux of each star in SDSS is monochromatic, so the narrow-band filter properties were applied to the flux to represent the flux in that filter. This method is susceptible to variations in the colors of the stars used, and variations in the filter transmission, and is thus not as accurate as observing spectrophotometric standard stars. Both standard star fluxes and image fluxes were corrected for airmass. 

\section{H$\alpha$ Imaging Results}

The r-band, and continuum-subtracted H$\alpha$ images are shown in the Appendix (Figures \ref{Haplot_660} to \ref{Haplot_10288}) for the sample galaxies. The continuum-subtracted H$\alpha$ images are shown in two stretches to accentuate both faint and bright structures. 

An H$\alpha$ flux is derived by summing each image, with care taken to subtract the minor excess sky background (where present) and exclude foreground stars when possible. For images that contain image artifacts, such as poorly continuum-subtracted foreground stars, the total flux was measured in images where these artifacts were replaced with noise typical to the image background for that frame. Due to the small wavelength coverage of the narrow-band H$\alpha$ filters, {\NII} contamination is generally low for the sample (see Table \ref{obstable}), and thus we do not correct for {\NII} contamination. The integrated H$\alpha$ results are included in Table \ref{sfrtable}, also with the $22$ micron results from \citet{wiegertetal15}, used in the following section to estimate SFR. Literature H$\alpha$ imaging was used to estimate the H$\alpha$ flux in the cases where no APO imaging was taken for this study (exceptions: NGC 4438 and NGC 5084). The literature observations used were from \citet{randetal92}, \citet{rand96}, \citet{collinsetal00}, the Spitzer Infrared Nearby Galaxy Survey (SINGS; \citealt{kennicuttetal03}), and the Spitzer Local Volume Legacy Survey (LVL; \citealt{daleetal09}), and are noted in the table caption.

Uncertainties in the H$\alpha$ flux for both the APO observed maps and literature maps take background variations and continuum subtraction uncertainties into account. Uncertainties in the revised SFR (see below) take H$\alpha$ flux uncertainties, $22$ micron image background uncertainties, and the SFR calibration uncertainties into account. Uncertainties in the galaxy distance are not considered in this analysis.  

\subsection{Star Formation Rates}

Empirical relationships between observed H$\alpha$ emission and mid-IR emission to extinction-corrected H$\alpha$ emission from \citet{calzettietal07} were employed:
\begin{equation}
\rm{L}(\rm{H}\alpha_{\rm{corr}}) = \rm{L}(\rm{H}\alpha_{\rm{obs}})+\rm{a}\cdot \nu \rm{L}_{\nu}(24\mu \rm{m})
\end{equation}

The H$\alpha$ measurements from the present study are used for the H$\alpha$ component to this relationship, and the WISE $22$ micron measurements from \citet{wiegertetal15} are used for the $24$ micron component (see the discussion on the equivalence of $22$ micron and $24$ micron emission below). 
Originally, a$=0.031$ was found in \citet{calzettietal07}. We adopt a$=0.042$ as an additional mid-IR extinction correction, as prescribed in \citet{vargasetal18} for use in edge-on or extremely dusty galaxies (reflecting an excess 22 micron attenuation of 1.36 in edge-on relative to non-edge-on galaxies). Applying the new correction to this edge-on sample is an improvement on past SFR estimates. This single-value mid-IR extinction correction was found by comparing integrated 25 micron to 100 micron flux ratios for a larger sample of galaxies. Thus, variations in the average dust content of star forming edge-on galaxies should be roughly accounted for using this method. A full discussion on the use of a single-value mid-IR extinction correction is included in \citet{vargasetal18}. We also note that these calibrations were made using $24$ micron flux, rather than the $22$ micron measurements we employ. Fluxes in the two slightly different mid-IR bands were found to have a tight linear correlation, and only differ by a factor of 1.03 \citep{wiegertetal15}. Thus, when $24$ micron fluxes are needed, we multiply our $22$ micron flux measurements by a factor of 1.03 to account for their subtle difference.

We then relate the extinction-corrected H$\alpha$ emission to SFR using the following relation from \citet{murphyetal11}:
\begin{equation}
\left( \frac{\rm{SFR}_{\rm{mix}}}{\rm{M_{\odot}}\rm{yr}^{-1}} \right) = 5.37\times 10^{-42} \left( \frac{\rm{L(\rm{H\alpha_{obs}})} +a\cdot \nu \rm{L}_{\nu}(24 \mu \rm{m})}{\rm{erg} \cdot \rm{s}^{-1}}\right)
\end{equation}

\begin{table*}[h]
\addcontentsline{lot}{table}{Integrated H$\alpha$ Results and SFR}
\tiny
\centering

    \begin{tabular}{c c c c c c c c}
    \hline
    \hline
    Galaxy & 
    D$^{\dagger}$ (Mpc) & 
    \hspace{-1cm}\thead{\tiny F$_{\rm{H}\alpha}$\\ \tiny ($\times10^{-12}$ erg s$^{-1}$ cm$^{-2}$)} &
     \hspace{-1cm}\thead{\tiny L$_{\rm{H}\alpha}$\\ \tiny ($\times10^{41}$ erg s$^{-1}$)} & 
    \hspace{-0.85cm} \thead{\tiny L$_{22\mu \rm{m}}^{\dagger}$ \\ \tiny ($\times10^{41}$ erg s$^{-1}$) } &
     \hspace{-0.85cm} \thead{\tiny SFR$_{22\mu \rm{m}}^{\dagger}$ \\ \tiny (M$_{\odot}$ yr$^{-1}$)} &
      \hspace{-0.85cm}\thead{\tiny SFR$_{\rm{revised}}$ \\ \tiny (M$_{\odot}$ yr$^{-1}$)} & 
       \hspace{-0.85cm} \thead{\tiny SFR$_{\rm{SD}}$ \\ \tiny ($\times10^{-3}$ M$_{\odot}$ yr$^{-1}$ kpc$^{-2}$)}\\
    \hline
    NGC 660  & 12.3 & 1.39$\pm$0.29 & 0.25$\pm$0.05 & 140.7$\pm$38.0 & 2.74$\pm$0.69  & 3.31$\pm$0.32 & 36.1$\pm$3.1 \\ 
    NGC 891  & 9.1 & 1.61$\pm$0.24 & 0.16$\pm$0.02  & 79.7$\pm$21.5  & 1.55$\pm$0.39  & 1.88$\pm$0.18  & 3.81$\pm$0.36\\
    NGC 2613 & 23.4 & 3.88$\pm$0.80 & 2.54$\pm$0.52 & 88.5$\pm$23.9 & 1.73$\pm$0.43   & 3.36$\pm$0.35  & 3.77$\pm$0.39\\
    NGC 2683 & 6.27$^*$ & 5.47$\pm$1.00 & 0.25$\pm$0.05  & 4.75$\pm$1.28 & 0.09$\pm$0.02  & 0.25$\pm$0.03  & 3.54$\pm$0.40\\
    NGC 2820 & 26.5 & 1.39$\pm$0.28 & 1.17$\pm$0.23 & 31.9$\pm$8.62 & 0.62$\pm$0.16    & 1.35$\pm$0.14  & 9.00$\pm$0.96\\
    NGC 2992 & 34 & 2.96$\pm$0.52 & 4.09$\pm$0.71   & 164.6$\pm$44.4 & 3.22$\pm$0.81   & 5.91$\pm$0.54 & 48.2$\pm$4.4\\
    NGC 3003 & 25.4 & 1.87$\pm$0.34 & 1.44$\pm$0.26 & 34.7$\pm$9.38 & 0.67$\pm$0.17   & 1.56$\pm$0.16  & 2.59$\pm$0.27\\
    NGC 3044 & 20.3 & 2.44$\pm$0.41 & 1.20$\pm$0.20 & 48.8$\pm$13.2 & 0.95$\pm$0.24   & 1.75$\pm$0.16  & 6.79$\pm$0.60\\
    NGC 3079 & 20.6 & 3.93$\pm$0.70 & 2.00$\pm$0.36 & 177.7$\pm$48.0 & 3.46$\pm$0.87  & 5.08$\pm$0.45 & 9.57$\pm$0.84\\
    NGC 3432 & 9.42$^*$ & 5.58$\pm$0.88 & 0.59$\pm$0.09 & 7.86$\pm$2.12 & 0.15$\pm$0.04  & 0.51$\pm$0.06  & 6.36$\pm$0.69\\
    NGC 3448 & 24.5 & 1.85$\pm$0.37 & 1.33$\pm$0.27 & 47.41$\pm$12.8 & 0.92$\pm$0.23     & 1.78$\pm$0.18 & 14.5$\pm$1.5\\
    NGC 3556 & 14.09 & 8.30$\pm$1.31 & 1.97$\pm$0.31 & 111.2$\pm$30.0 & 2.17$\pm$0.54    & 3.57$\pm$0.30 & 7.32$\pm$0.62\\
    NGC 3628 & 8.5 & 5.07$\pm$0.76 & 0.44$\pm$0.07 & 52.1$\pm$14.1 & 1.01$\pm$0.25     & 1.41$\pm$0.12  & 2.99$\pm$0.26\\
    NGC 3735 & 42  & 0.71$\pm$0.13 & 1.50$\pm$0.27 & 240.7$\pm$65.0 & 1.10$\pm$0.27    & 6.23$\pm$0.57 & 6.71$\pm$0.61\\
    NGC 3877 & 17.7 & 1.47$\pm$0.26 & 0.55$\pm$0.10 & 46.9$\pm$12.7 & 0.92$\pm$0.23    & 1.35$\pm$0.12  & 5.04$\pm$0.44\\
    NGC 4013 & 16   & 0.98$\pm$0.21 & 0.30$\pm$0.07 & 24.5$\pm$6.62 & 0.48$\pm$0.12    & 0.71$\pm$0.07  & 3.51$\pm$0.32\\
    NGC 4096 & 10.32& 5.88$\pm$1.15 & 0.75$\pm$0.14 & 13.8$\pm$3.72 & 0.27$\pm$0.07    & 0.71$\pm$0.08  & 6.52$\pm$0.77\\
    NGC 4157 & 15.6 & 2.00$\pm$0.65 & 0.58$\pm$0.19 & 64.4$\pm$17.4 & 1.25$\pm$0.31    & 1.76$\pm$0.18  & 8.15$\pm$0.83\\ 
    NGC 4192 & 13.55& 1.09$\pm$0.20 & 0.24$\pm$0.04 & 28.8$\pm$7.77 & 0.56$\pm$0.14    & 0.78$\pm$0.07  & 1.67$\pm$0.15\\
    NGC 4217 & 20.6 & 0.43$\pm$0.06$^a$ & 0.22$\pm$0.03 & 78.7$\pm$21.3 & 1.53$\pm$0.38   & 1.89$\pm$0.18 & 4.40$\pm$0.42\\
    NGC 4244 & 4.4  & 2.94$\pm$0.44$^d$ & 0.07$\pm$0.01 & 1.20$\pm$0.33    & 0.02$\pm$0.01 & 0.06$\pm$0.01 & 0.37$\pm$0.04\\
    NGC 4302 & 19.41& 1.30$\pm$0.20$^a$ & 0.58$\pm$0.09 & 27.1$\pm$7.30    & 0.53$\pm$0.13 & 0.92$\pm$0.08 & 2.52$\pm$0.21\\
    NGC 4388 & 16.6 & 1.17$\pm$0.18 & 0.39$\pm$0.06 &     97.9$\pm$26.4    & 0.07$\pm$0.48 & 2.42$\pm$0.23 & 25.0$\pm$2.3\\
    NGC 4438 & 10.39& N/A  & N/A  &              3.36$\pm$0.91    & 0.07$\pm$0.02  & N/A  &  N/A \\ 
    NGC 4565 & 11.9 & 1.14$\pm$0.17 & 0.19$\pm$0.03  &    38.1$\pm$10.3    & 0.74$\pm$0.18   & 0.96$\pm$0.09 & 0.94$\pm$0.09  \\
    NGC 4594 & 12.7 & 0.67$\pm$0.10$^c$ & 0.13$\pm$0.02 & 16.2$\pm$4.38    & 0.32$\pm$0.08   & 0.43$\pm$0.04 & 1.14$\pm$0.10\\
    NGC 4631 & 7.4  & 30.8$\pm$4.62$^c$ & 2.02$\pm$0.30 & 68.2$\pm$18.4    & 1.33$\pm$0.33   & 2.62$\pm$0.22 & 6.10$\pm$0.52\\
    NGC 4666 & 27.5$^*$ & 4.20$\pm$0.69 & 3.80$\pm$0.62 &     373.8$\pm$100.9    & 7.29$\pm$1.82  & 10.5 $\pm$0.92 & 12.8$\pm$1.1\\
    NGC 4845 & 16.98& 0.38$\pm$0.09 & 0.13$\pm$0.03 &     24.5$\pm$6.61    & 0.48$\pm$0.12  & 0.62$\pm$0.06 & 7.33$\pm$0.68\\
    NGC 5084 & 23.4 & N/A  & N/A  &              5.24$\pm$1.42    & 0.10$\pm$0.03  & N/A & N/A\\
    NGC 5297 & 40.4 & 1.47$\pm$0.28 & 2.88$\pm$0.54 &     64.5$\pm$17.4    & 1.27$\pm$0.32  & 3.00$\pm$0.33 & 5.70$\pm$0.62\\
    NGC 5775 & 28.9$^*$ & 0.38$\pm$0.41$^b$ & 2.70$\pm$0.41 & 270.9$\pm$73.1    & 5.28$\pm$1.32 & 7.56$\pm$0.65 & 9.40$\pm$0.81\\
    NGC 5792 & 31.7$^*$ & 2.13$\pm$0.32  & 2.56$\pm$0.38 &    134.7$\pm$36.4    & 2.63$\pm$0.66 & 4.41$\pm$0.37  & 10.0$\pm$0.8\\
    NGC 5907 & 16.8 & 1.47$\pm$0.34$^a$ & 0.77$\pm$0.12 & 79.7$\pm$21.5    & 1.56$\pm$0.39 & 2.21$\pm$0.19 & 2.18$\pm$0.19\\
    UGC 10288 & 34.1& 0.25$\pm$0.08 & 0.35$\pm$0.10 &     20.9$\pm$5.64    & 0.41$\pm$0.10 & 0.66$\pm$0.07 & 1.85$\pm$0.21\\   

    \hline   
    \end{tabular}
    
\caption{\small Integrated H$\alpha$ results and integrated $22$ micron fluxes for the CHANG-ES sample. The distances (D) column contains values quoted in \cite{wiegertetal15}, which have uncertainties of $\sim25\%$, except for values marked by the `$*$' symbol, where uncertainties are $\sim35\%$. The values of SFR$_{22\mu \rm{m}}$ were computed in \citet{wiegertetal15} using only the $22$  micron flux. SFR$_{\rm{revised}}$ values were found using a combination of H$\alpha$ and $22$ micron data, as outlined in \citet{vargasetal18}. SFR$_{\rm{SD}}$ values were computed from SFR$_{\rm{revised}}$ and the $22$ micron diameters from \cite{wiegertetal15}. \\
$\dagger~$values from \citet{wiegertetal15}. These values do not contain any mid-IR extinction correction. \\
$a~$values obtained using imaging from \citet{rand96}. \\
$b~$values obtained using imaging from \citet{collinsetal00}\\
$c~$values obtained using imaging from SINGS \\
$d~$values obtained using imaging from LVL  \\ 
}
\label{sfrtable}
\end{table*}

\begin{figure*}
\centering
\includegraphics[scale=0.65]{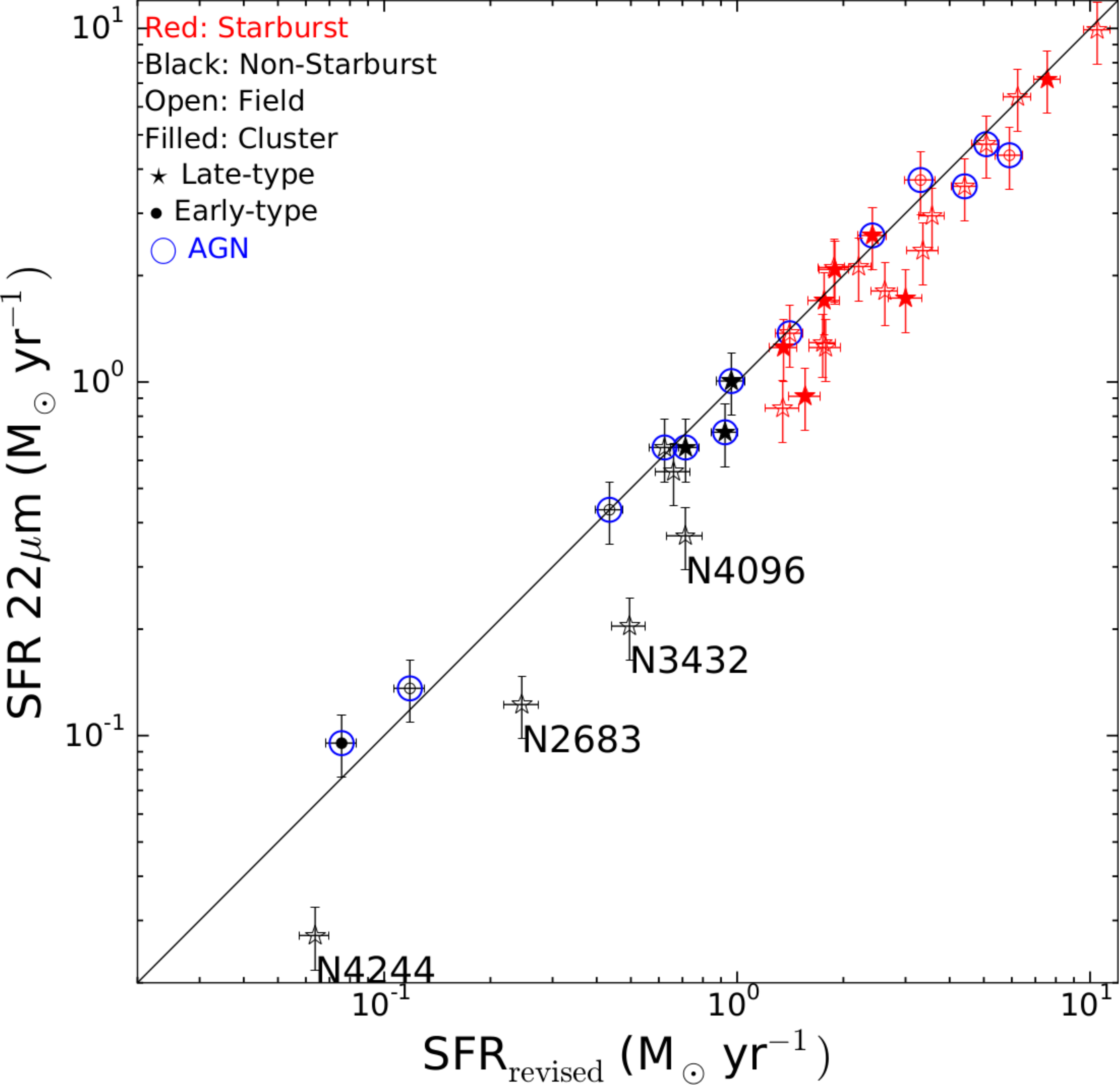}
\caption[SFR Comparison with Updated Method]{Comparison of SFR taken from the CHANG-ES D Configuration data release paper, \citet{wiegertetal15}, and SFR from this study. The SFRs from this study are obtained using the revised method with $a=0.042$, and the SFR from \citet{wiegertetal15} use the 22 micron only calibration from \citet{jarrettetal13}. A mid-IR extinction correction of factor 1.36 has been added to the SFR estimates from \citet{wiegertetal15}. The data points use color coding and symbols to denote galaxy properties. The solid line is a one-to-one line, and outliers are labelled. }
\label{sfr_comp_136_plot}
\end{figure*}

The SFR estimates derived using only WISE $22$ micron flux in \citet{wiegertetal15} are systematically lower than those of this work, which combine $22$ micron flux with H$\alpha$ flux. This is due to the nature of the SFR calibrations and to the addition of the mid-IR extinction correction for edge-on galaxies discussed in \citet{vargasetal18}. For example, in a hypothetical galaxy with zero H$\alpha$ emission, the original H$\alpha$ $+$ mid-IR SFR would be almost as large as the mid-IR only SFR (\ie SFR$_{\rm{mix}}=0.83~\cdot$ SFR$_{22\mu\rm{m}}$, for zero H$\alpha$ emission). Hence, a moderately star-forming galaxy's H$\alpha$ emission will cause SFR$_{\rm{mix}}$ to exceed SFR$_{22\mu\rm{m}}$. Furthermore, we add the aforementioned edge-on mid-IR extinction correction factor in our new H$\alpha$ $+$ mid-IR SFR estimates, referred to here as SFR$_{\rm{revised}}$, which increases the SFR estimates from previous uncorrected (and likely underestimated) versions. The old and new SFR values are brought into better agreement by increasing the \citet{wiegertetal15} SFRs by the aforementioned attenuation factor of 1.36. In Figure \ref{sfr_comp_136_plot}, we plot these values against the SFR$_{\rm{revised}}$ values from this work. Even with the new extinction correction applied to both, one cannot expect the two estimates to be in excellent agreement because of slight differences in the original studies upon which they are based, namely \citet{riekeetal09} for the 22 micron estimate and \citet{calzettietal07} for the mid-IR and H$\alpha$ mixture method.

We also analyze the SFR per unit area, or the SFR surface density (SFR$_{\rm{SD}}$). We use the WISE $22$ micron diameter values estimated in \citet{wiegertetal15} to calculate SFR$_{\rm{SD}}$. We note that SFR$_{\rm{SD}}$ is the same quantity as $\Sigma_{\rm{SFR}}$, but with nomenclature selected for consistency with past CHANG-ES publications. 

Different symbols and colors are plotted for each galaxy in Figure \ref{sfr_comp_136_plot} and subsequent sample figures, based on their physical properties. For consistency, we use the same color and symbol scheme as the figures in \citet{lietal16} (CHANG-ES Paper VI). Also consistent with \citet{lietal16}, we use the same definition of a starburst galaxy (SFR $>1$ M$_{\odot}$ yr$^{-1}$ and SFR$_{\rm{SD}}$ $>0.002$  M$_{\odot}$ yr$^{-1}$ kpc$^{-2}$, using the updated SFR and SFR$_{\rm{SD}}$ values from this work. We also use the same definitions for field vs. cluster galaxies, late-type vs. early-type, and AGN activity as outlined in \citet{lietal16}.  

There are four notable outliers in Figure \ref{sfr_comp_136_plot}: NGC 2683, NGC 3432, NGC 4096, and NGC 4244. Each of these are late-type field galaxies with relatively low SFR. It is likely that these galaxies host  characteristically different dust populations than the mean galaxy in the RBGS analysis done in \citet{vargasetal18}, which would cause them to lay further off the one-to-one line in Figure \ref{sfr_comp_136_plot}. An individual analysis on the dust content of each outlier would be needed to determine if a different dust extinction correction is appropriate, on a case-by-case basis.  

\section{Comparison to Radio Scale Height}

The radio continuum scale heights of a subset of CHANG-ES were calculated in \citet{krauseetal18}. The analyzed subsample consists of NGC 2820, NGC 3003, NGC 3044, NGC 3079, NGC 3432, NGC 3735, NGC 3877, NGC 4013, NGC 4157, NGC 4217, NGC 4302, NGC 5775, and UGC 10288. 
We analyze the correlation between radio scale heights for the subset of CHANG-ES galaxies in \citet{krauseetal18} and improved SFR estimates from this study. As in \citet{lietal16}, we use the Spearman's rank correlation coefficient ($r_s$) to quantitatively assess the presence and strength of correlations. We also adopt the same definitions for correlation strength as in \citet{lietal16}: $|r_s| \geq 0.6$ is a strong correlation, $0.3 \leq |r_s| < 0.6$ is a weak correlation, and $|r_s|<0.3$  is no correlation. We present the Spearman's rank correlation coefficients for all considered quantities in Table \ref{correlationtable}. We also include the logarithmic slope of a simple linear fit to the data, which may be of interest to future studies.

\begin{table*}[h]
\begin{center}
  \begin{tabular}{cccccc}
\hline
\hline
 &\multicolumn{2}{c}{SH}
&&
\multicolumn{2}{c}{NSH} \\\cline{2-3}\cline{5-6}
 &$r_s$& $m$    && $r_s$ & $m$      \\
 SFR ~(C-band) & 0.75 & 0.28 && -0.10 & N/A\\
 SFR ~(L-band) & 0.83 & 0.27 && -0.19 & N/A\\
 SFR$_{\rm{SD}}$ (C-band) & 0.55 & 0.36 && 0.36 & 0.21 \\
 SFR$_{\rm{SD}}$ (L-band) & 0.85 & 0.26 && 0.52 & 0.30  \\ 
\hline
\hline

\end{tabular}
\end{center}    
\caption{\small Spearman's rank correlation coefficients, $r_s$, and logarithmic slope of a linear fit to the data, $m$, for SFR and SFR$_{\rm{SD}}$ vs. radio scale height (SH) and normalized scale height (NSH) values in both C-band and L-band from \citet{krauseetal18}. All correlations are estimated with NGC 3003 and NGC 3079 removed. Correlations are defined as $|r_s| \geq 0.6$ for a strong correlation, $0.3 \leq |r_s| < 0.6$ for a weak correlation, and $|r_s|<0.3$ for no correlation. Uncertainties in correlation coefficients are of order $\sim0.1$. Slope values are only given for quantities with at least weak correlation.}
\label{correlationtable}
\end{table*}

The scale height estimate of NGC 3003 is a clear outlier. \citet{krauseetal18} remark that the inclination of NGC 3003 is not well-constrained, and it seems to be disturbed due to a recent interaction. Thus, the scale height estimate for NGC 3003 is likely unreliable. We removed NGC 3003 for this analysis, and present only results excluding NGC 3003. As discussed in \citet{krauseetal18}, we also note that the scale height of NGC 3079 may be underestimated due to the poor quality of the L-band map. This could explain its unphysical L-to-C-band scale height ratio, which should be $>1$. With this in mind, we also exclude NGC 3079 from our analysis. The change in the correlation coefficients before and after removing the single galaxy NGC 3079 was at most $\sim0.1$. Thus, we use this as a rough estimate for the uncertainty of the correlation coefficients. 

\begin{figure*}
\centering
\includegraphics[scale=0.50]{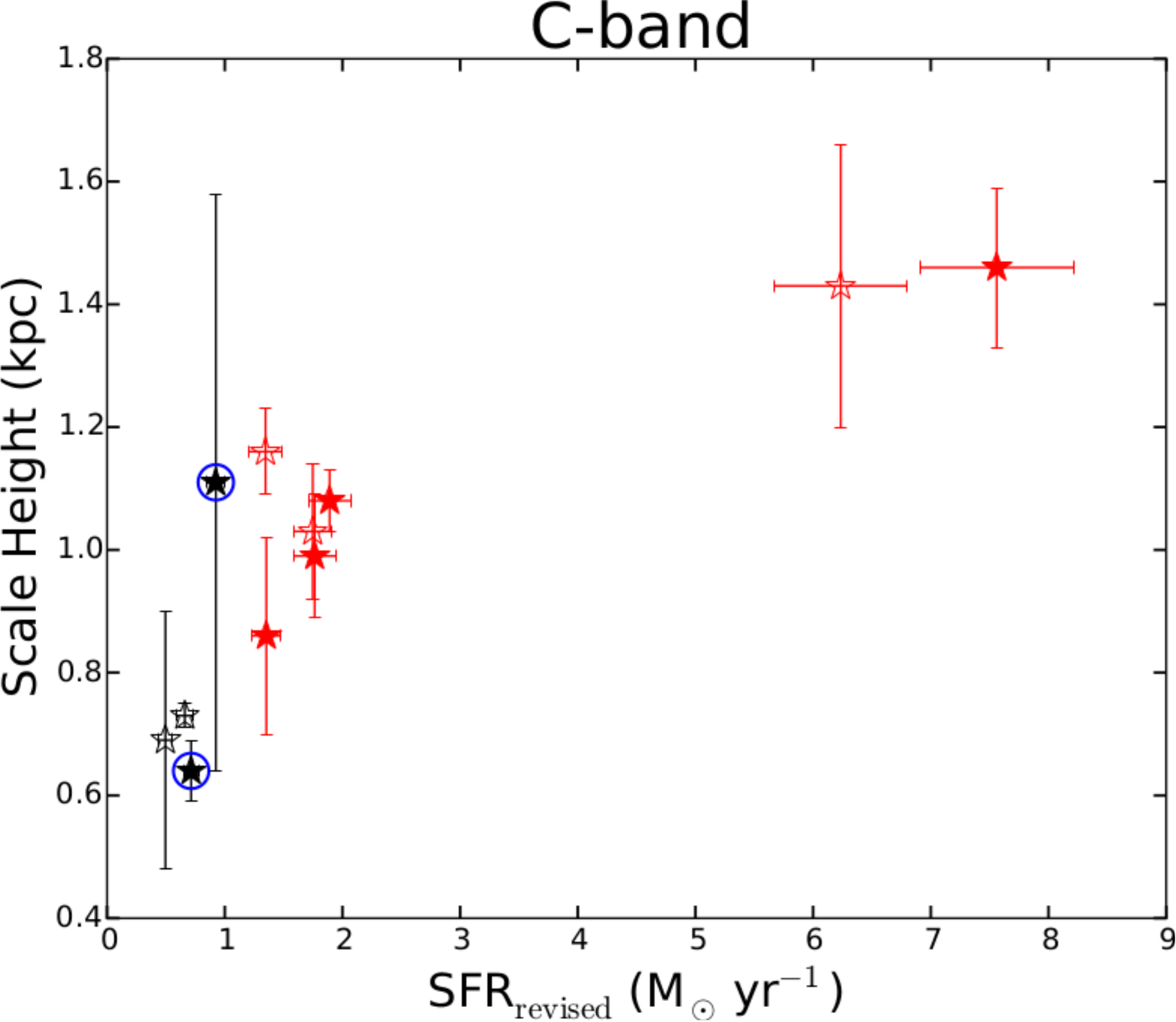}
\includegraphics[scale=0.45]{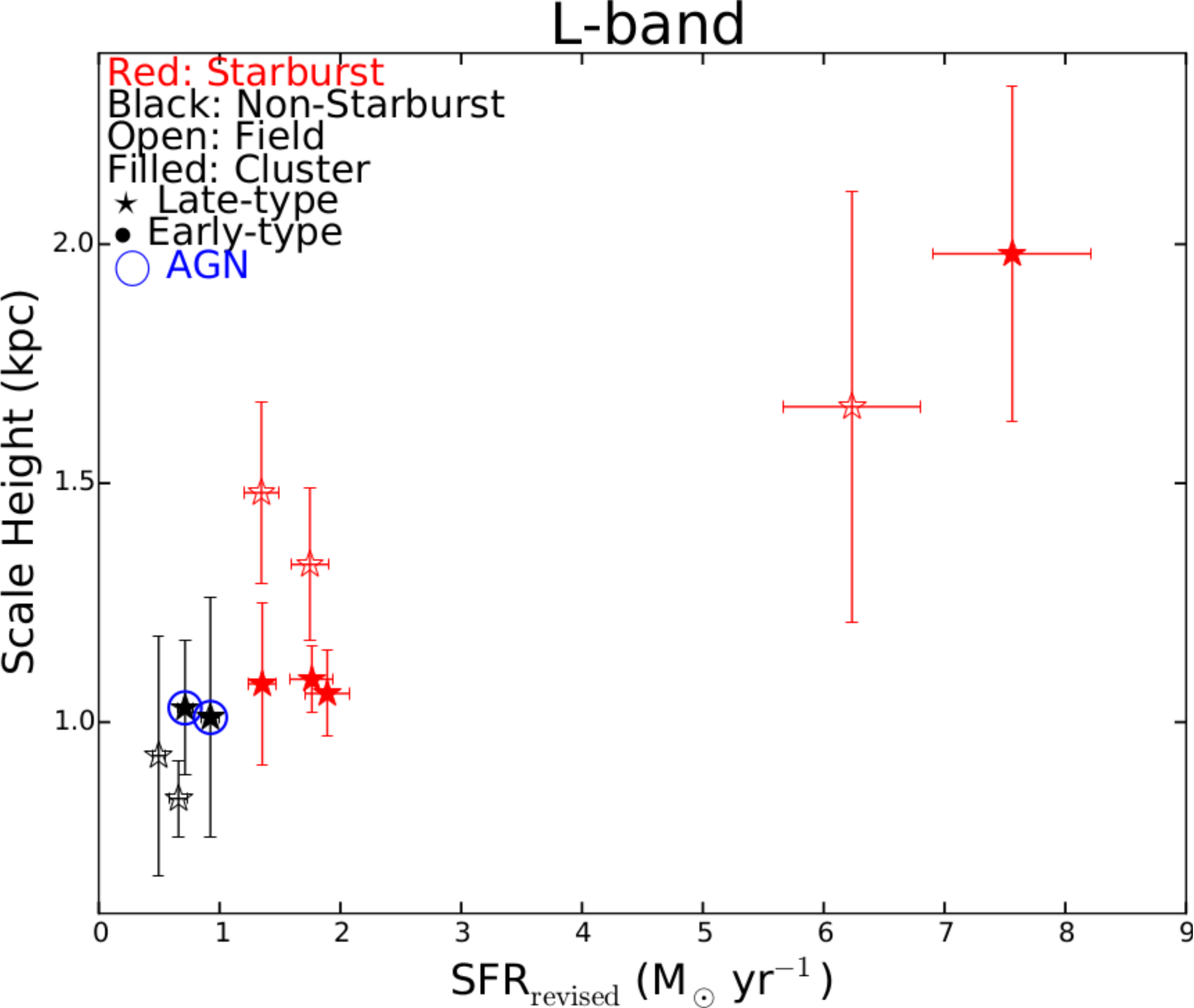}
\caption[Scale Height vs. SFR with NGC 3003 Removed]{C-band (top) and L-band (bottom) scale height values from \citet{krauseetal18} versus SFR from this work with NGC 3003 and NGC 3079 removed. The color and symbol scheme is the same as Figure \ref{sfr_comp_136_plot}. }
\label{SH_SFR_no3003}
\end{figure*}

\begin{figure*}
\centering
\includegraphics[scale=0.45]{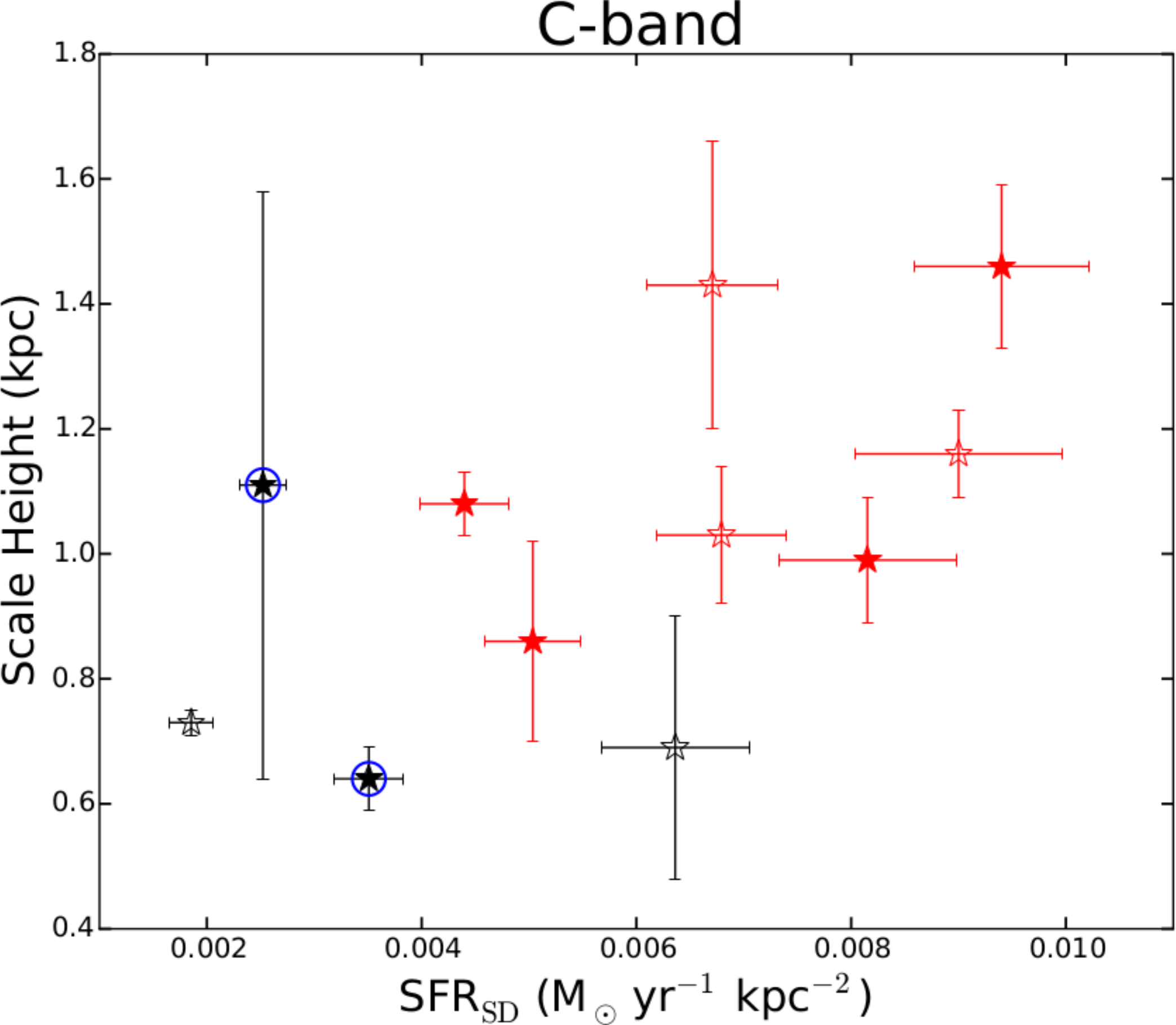}
\includegraphics[scale=0.45]{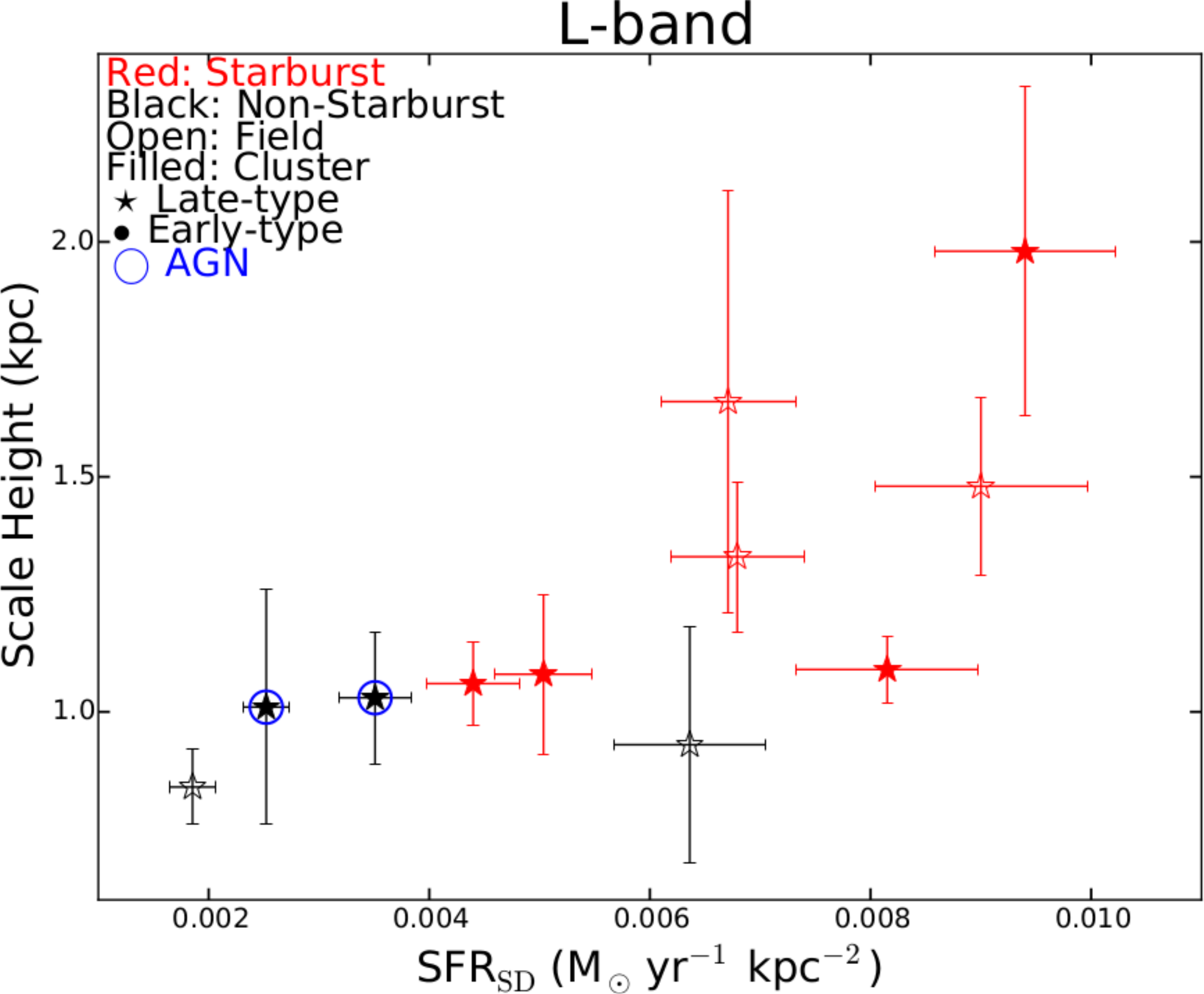}
\caption[Scale Height vs. SSFR with NGC 3003 Removed]{C-band (top) and L-band (bottom) scale height values from \citet{krauseetal18} versus SFR$_{\rm{SD}}$ from this work with NGC 3003 and NGC 3079 removed. The color and symbol scheme is the same as Figure \ref{sfr_comp_136_plot}. }
\label{SH_SSFR_no3003}
\end{figure*}

Figure \ref{SH_SFR_no3003} shows the scale heights plotted against SFR, with NGC 3003 removed. The Spearman's rank correlation coefficient for these quantities indicates strong positive correlation. A similar, though slightly weaker correlation is seen with scale height and SFR$_{\rm{SD}}$, plotted in Figure \ref{SH_SSFR_no3003}. 

\begin{figure*}
\centering
\includegraphics[scale=0.45]{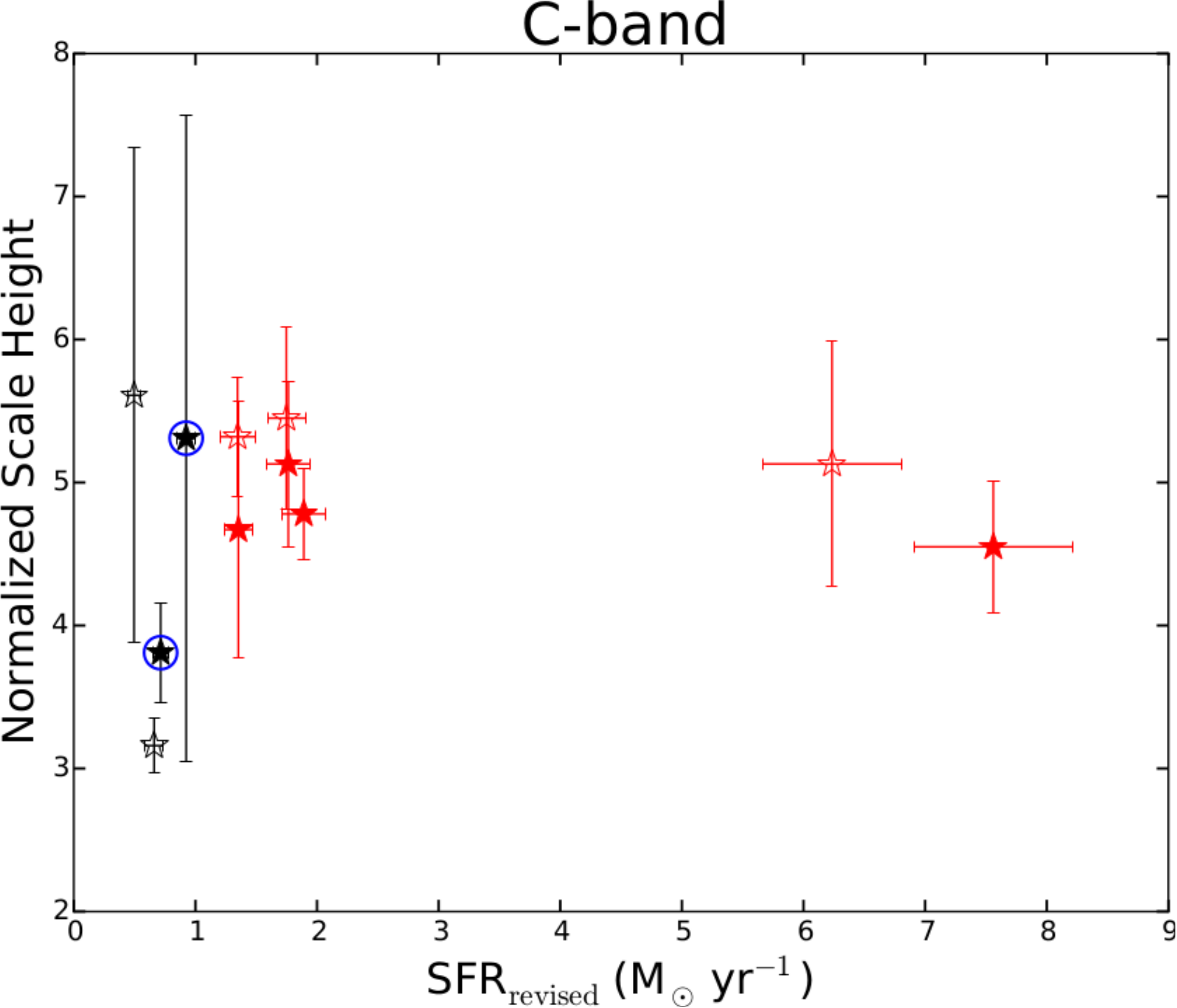}
\includegraphics[scale=0.45]{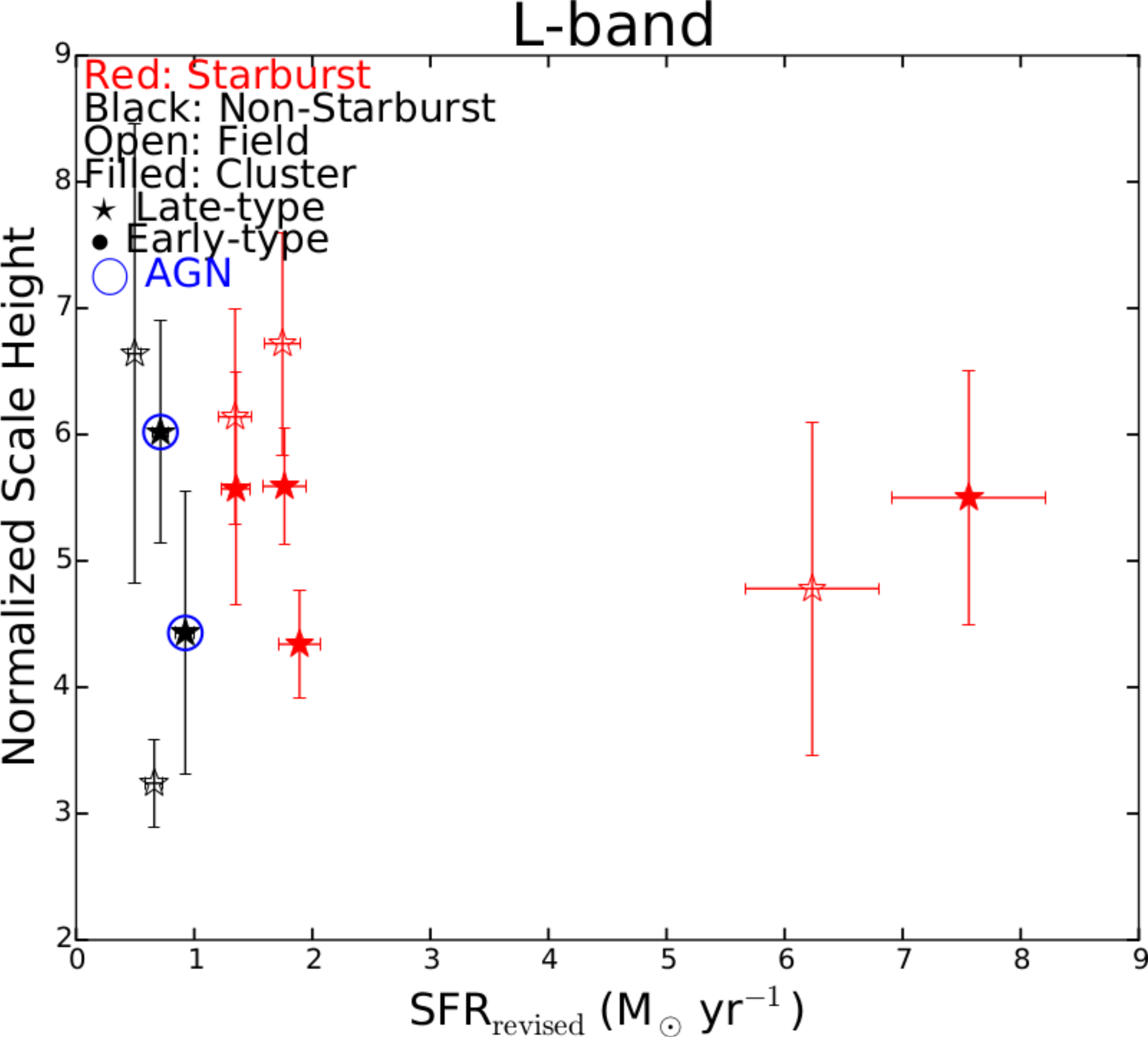}
\caption[Normalized Scale Height vs. SFR with NGC 3003 Removed]{C-band (top) and L-band (bottom) normalized scale height values from \citet{krauseetal18} versus SFR from this work with NGC 3003 removed. The color and symbol scheme is the same as Figure \ref{sfr_comp_136_plot}.}
\label{NSH_SFR_no3003}
\end{figure*}

\begin{figure*}
\centering
\includegraphics[scale=0.45]{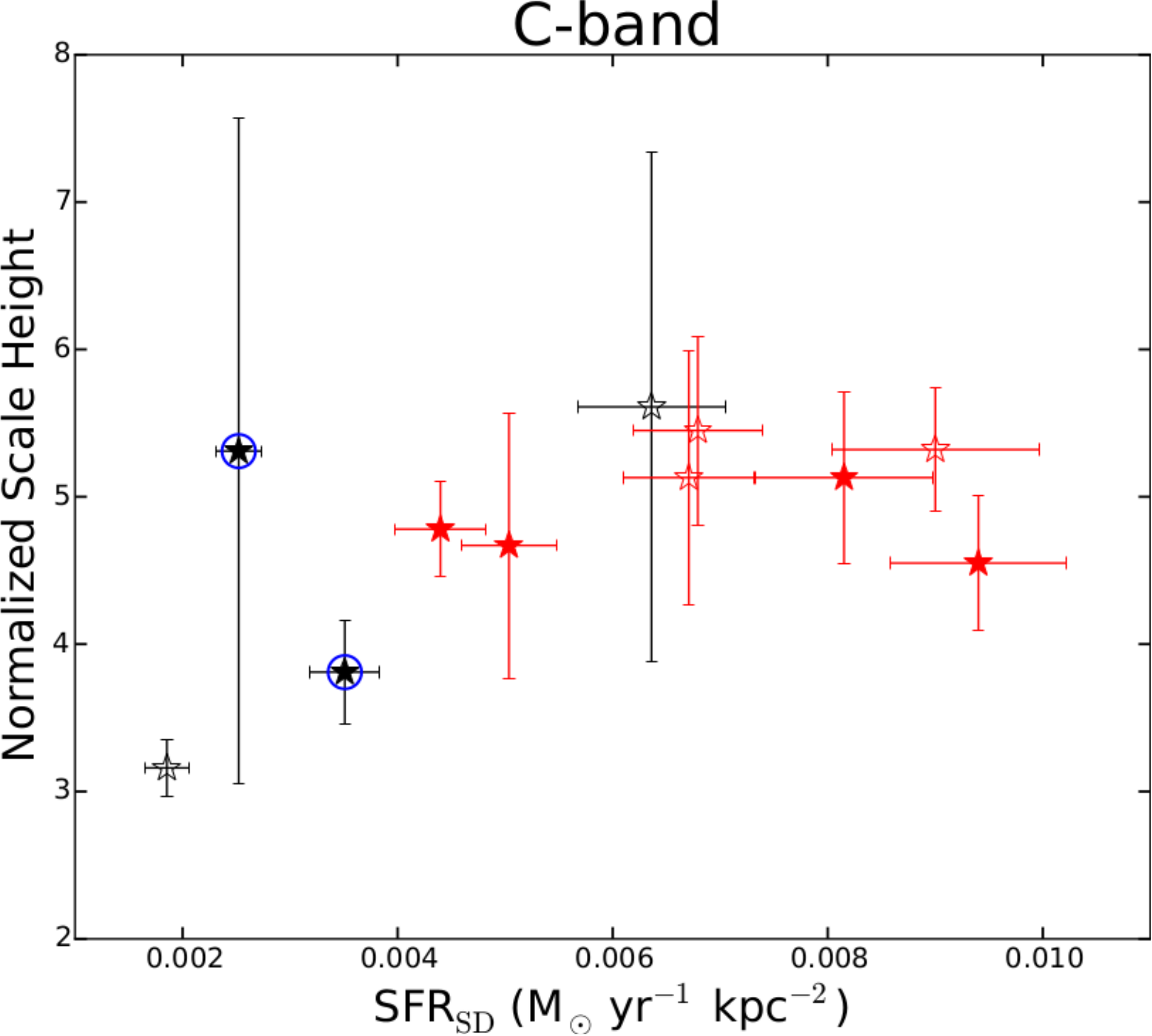}
\includegraphics[scale=0.45]{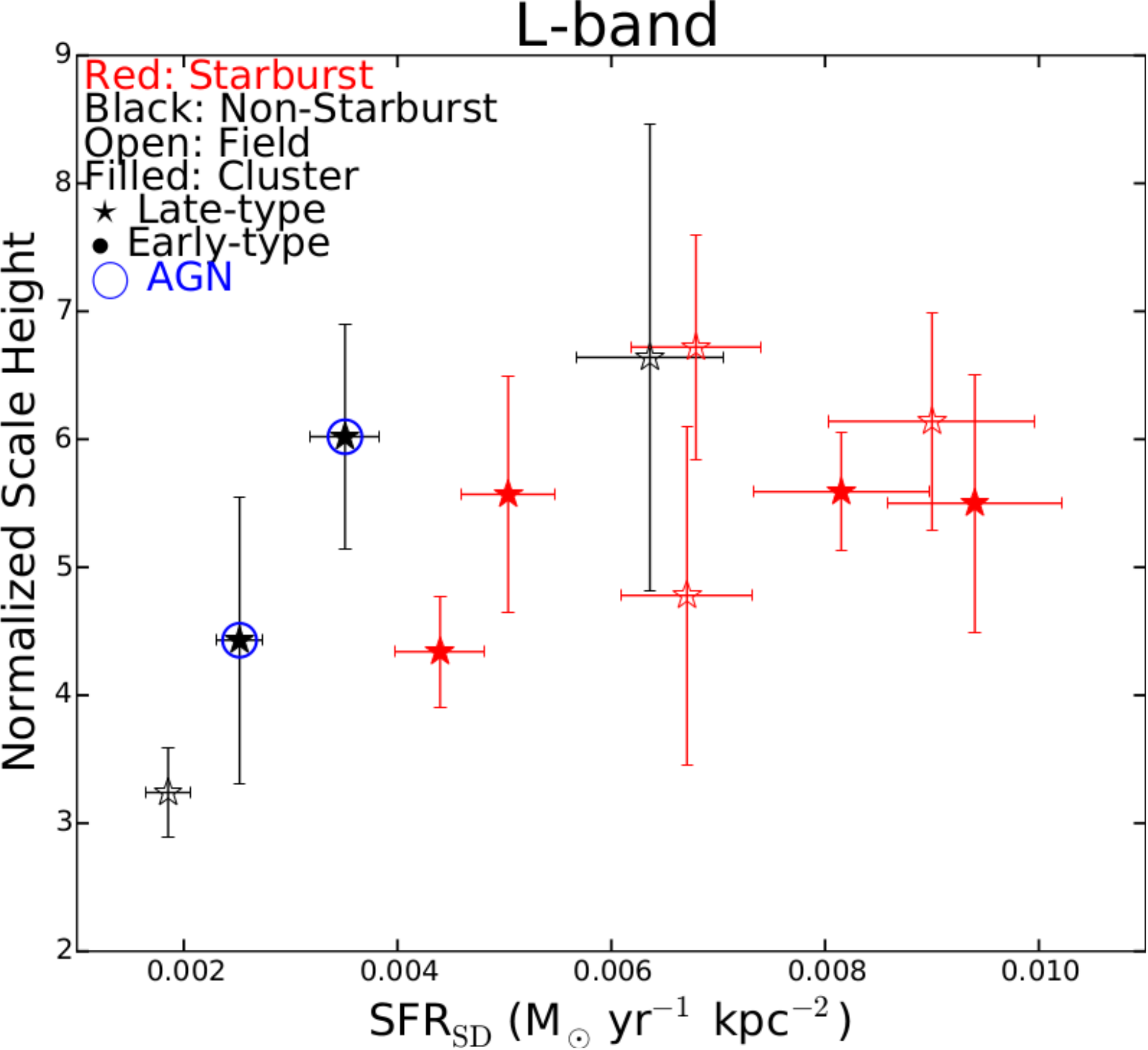}
\caption[Normalized Scale Height vs. SFR with NGC 3003 Removed]{C-band (top) and L-band (bottom) normalized scale height values from \citet{krauseetal18} versus SFR$_{\rm{SD}}$ from this work with NGC 3003 removed. The color and symbol scheme is the same as Figure \ref{sfr_comp_136_plot}.}
\label{NSH_SFRSD_no3003}
\end{figure*}

We also consider SFR relationships with the normalized scale height (NSH) estimates, also from \citet{krauseetal18}. NSH is defined as $100 \cdot h/d_r$, where $h$ is the radio scale height, and $d_r$ is the radio diameter at the same frequency. We plot SFR vs NSH in Figure \ref{NSH_SFR_no3003} and Figure \ref{NSH_SFRSD_no3003}. The Spearman's rank correlation coefficient shows no correlation between SFR and NSH, and weak correlation between SFR$_{\rm{SD}}$ and NSH in both radio bands.

No strong correlation is found between SFR or SFR$_{\rm{SD}}$ and NSH, which is in agreement with the results of \citet{krauseetal18}. However, \citet{krauseetal18} rejected this correlation in their data due to high $\chi^2$ values of their linear fit -- 2.0 in L-band and 5.5 in C-band \citep{krauseprivatecomm}. For a more direct comparison, we estimate the Spearman's rank correlation coefficient for SFR vs. radio scale height using the same SFR estimates as \citet{krauseetal18}, and also including both NGC 3003 and NGC 3079. We find $r_s=0.60$ in L-band and $r_s=0.69$ in C-band. While this implies there may originally have been some correlation, the correlation is no doubt strengthened in this present work. We attribute this to the updated SFR estimates of this study, which include H$\alpha$ information and the mid-IR extinction correction from \citet{vargasetal18}. The galaxies with the largest change in their updated SFR were the ones with the most contribution from H$\alpha$ in their revised SFR values. This, in conjunction with the new SFR -- radio scale height correlation, suggests that mid-IR extinction correction may be essential when estimating SFRs in edge-on galaxies. Previous estimates of SFR in edge-on galaxies using mid-IR only calibrations without a mid-IR extinction correction may be inaccurate. We also note that variations in the two SFR estimates is expected, as they are derived from fundamentally different empirical SFR calibrations. 

Interestingly, the correlation between SFR and and C-band radio scale height is slightly stronger than that between SFR$_{\rm{SD}}$ and radio scale height. This would imply that radio scale heights are not strictly governed by the underlying star formation, as that would produce a stronger correlation with SFR$_{\rm{SD}}$. Radio scale heights should depend on both the behavior of the magnetic field and the availability of cosmic ray electrons (CREs). Since CREs are accelerated and injected into the ISM by star formation processes, the availability of CREs is tied into SFR$_{\rm{SD}}$. The behavior of the magnetic field likely depends on the mechanism of the magnetic dynamo on all scales and possibly perturbations from regions of star formation. However, \citet{moraetal13} find evidence for a global regularity of the magnetic field in NGC 4631 on scales of several kpc. Since regions of star formation occur on smaller scales, a `disentanglement' of the small-scale magnetic field and the underlying star formation regions may exist above a certain size scale threshold. A weakened correlation between SFR$_{\rm{SD}}$ and radio scale height -- which in this case was measured on scales larger than typical {\HII} regions --  can be explained through the above scenario. SFR$_{\rm{SD}}$ has larger uncertainties than SFR due to its added dependence on galaxy size, so some caution is required when interpreting this correlation.     

\citet{krauseetal18} also find a strong linear correlation between radio scale height and radio diameter. Those authors state, ``We could not find an obvious physical explanation that would not simultaneously imply a correlation of the halo scale height with SFR or SFR$_{\rm{SD}}$." This intuitively flows from the idea that a larger star-forming galaxy would have more regions of star formation, and thus a larger SFR, simply due to its size. So, it is natural to see larger radio scale heights in larger galaxies, as we now see the correlation between scale height and both SFR and SFR$_{\rm{SD}}$ in the present study. 

\begin{center}
\begin{figure*}

\includegraphics[scale=0.55]{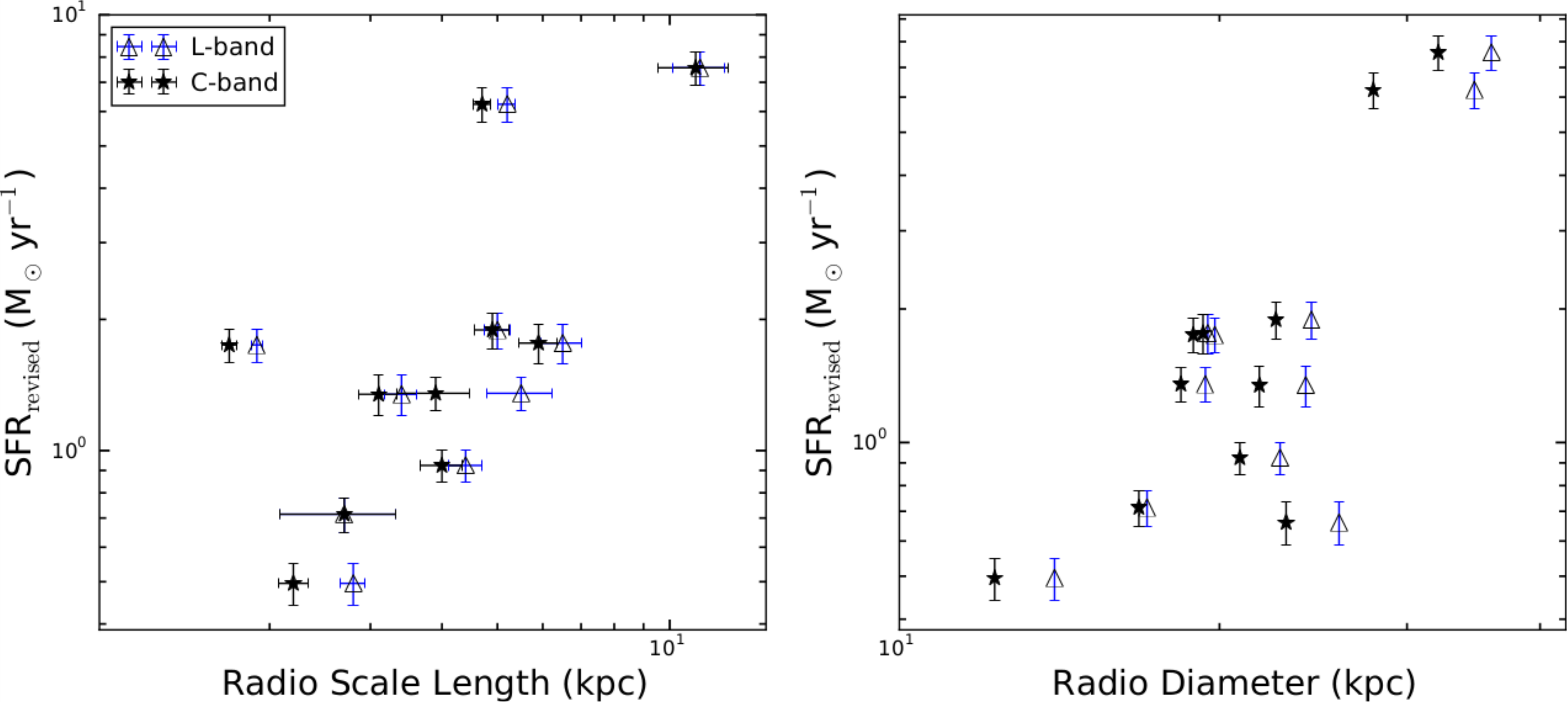}
\caption{Revised SFR vs. radio scale length values from \citet{krauseetal18} (left) and revised SFR vs. radio diameter (right). C-band points are plotted with black stars, and L-band points are plotted with blue open triangles. }
\label{SFR_lengthDiam}
\end{figure*}
\end{center}

As a further diagnostic into the origins of the radio scale height -- diameter relation, we plot SFR against radio scale length and radio diameter in Figure \ref{SFR_lengthDiam}. We find strong correlation in both SFR vs. radio scale length (C-band $r_s=0.78$; logarithmic slope $=0.43$) and SFR vs. radio diameter (C-band $r_s=0.61$; logarithmic slope $=0.24$). The existence of these SFR correlations with galaxy size indicate that the radio scale height -- diameter relation originates from star formation properties within the disk.   

\citet{leslieetal17}, a study of 6 edge-on galaxies in the SAMI survey, claim a connection was found between vertical radio extent and SFR in three of their galaxies. Though we note vertical extent depends on the sensitivity of observations, we build on those results by including scale height measurements. Future Optical IFU studies of the high SFR, high radio scale height galaxies in CHANG-ES may find evidence for small scale vertical outflows from the disk, bridging the link between cosmic ray injection via disk star formation processes and extended radio continuum emission from the inner halos of galaxies

A future study is planned that will include thermal radio continuum component prediction maps utilizing the released H$\alpha$ imaging from this present work. These refined estimates of thermal emission would likely have an affect on the results of \citet{krauseetal18}, namely their magnetic field estimates. In \citet{krauseetal18}, the thermal radio component was assumed to be a constant $20\%$ of the total in C-band and negligible in L-band. In \citet{vargasetal18}, large variations in thermal fraction both within a galaxy's disk and on a galaxy-to-galaxy basis were found. In the most extreme case, NGC 3044 showed C-band thermal fractions of $\sim40\%$, increasing up to $\sim80\%$ in certain discrete disk regions. It is possible that the `true' thermal fractions are larger than the estimated $20\%$ in C-band and non-negligible in L-band. The affect this would have on the average magnetic field strength is unclear however, since the field strength depends on both the non-thermal intensity and the non-thermal spectral index. With a higher thermal fraction, the non-thermal spectral index will steepen, acting to increase the field strength. However, the decreased non-thermal intensity would decrease the field strength. Since these effects act against each other, it is difficult to determine the net change in field strength without performing the analysis. We also note that the the scale height measurements from \citet{krauseetal18} would not be affected, since they represent the total radio continuum emission scale height, and not specifically the non-thermal (or thermal) component.  

\section{Extreme Outer Disk Star Formation and the XUV disk of NGC 4157}

\begin{figure*}
\centering
\includegraphics[scale=0.47]{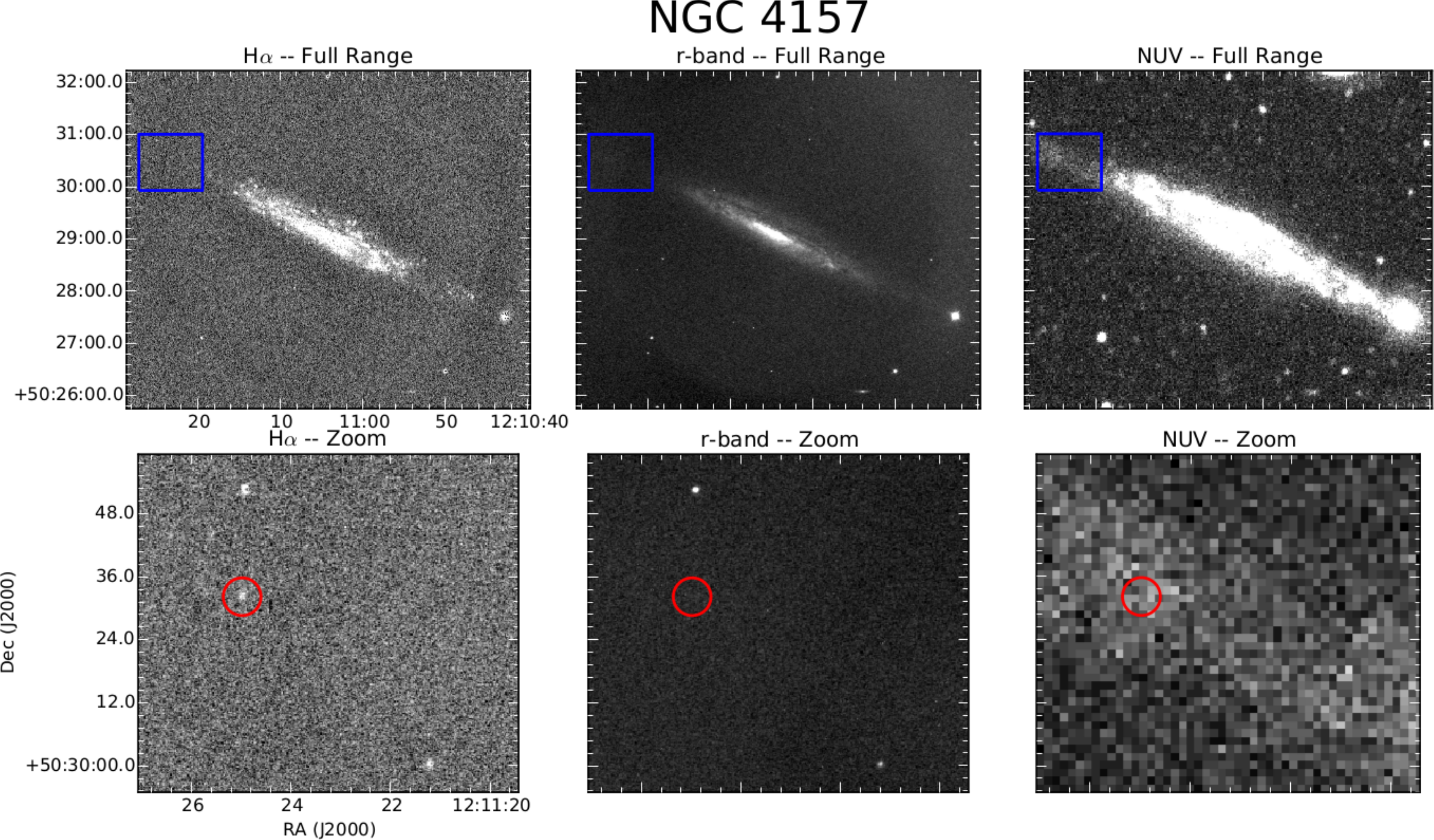}
\caption{XUV disk structure of NGC 4157. Upper row from left to right: H$\alpha$ narrow-band image, r-band continuum image, and GALEX NUV image of the full disk of NGC 4157. Lower row from left to right: H$\alpha$ narrow-band image, r-band continuum, and GALEX NUV image of the zoomed-in region associated with the outer disk star formation feature and XUV disk. The blue boxes in the upper row images outline the display regions of the zoomed-in panels. The red circles highlight the location of the extreme outer disk star formation feature. Note the existence of two imperfectly-subtracted foreground stars visible in both the H$\alpha$ and r-band zoomed-in panels.}
\label{XUV_4157}
\end{figure*}

We identified a source in the narrow-band H$\alpha$ image of NGC 4157 located well outside of the star forming disk, centered at $\alpha = 12^h11^m24.^s996$, $\delta=+50^{\circ}30^m32^s.04$. The source is located $\sim 60\arcsec$ ($\sim4.5$ kpc)  from the next nearest {\HII} region associated with the galaxy disk. The source shows no counterpart in the corresponding r-band image, thus it is likely a region of active star formation associated with NGC 4157. Interestingly, the region is spatially coincident with an extended ultraviolet (XUV) disk feature. We include an atlas of the feature in H-alpha, r-band, and GALEX NUV in Figure \ref{XUV_4157}. 

We attempt to constrain the ionizing population of the {\HII} region. The EM of the region is measured to be $\sim639$ pc cm$^{-6}$ from the H$\alpha$ image. The region is assumed to have a filling factor of unity within the volume implied by the H$\alpha$ image's seeing-limited FWHM of $\sim1\arcsec$. The C-band thermal radio flux density was estimated directly from the region's EM, and was found to be $\sim1.6$ $\mu$Jy (below the CHANG-ES sensitivity). This assumes T$_e=8000$ K, and is corrected by a factor of $1\arcsec /(15\arcsec \times 15\arcsec)$ for the incomplete filling factor within the radio beam. From this, we estimate the ionizing photon luminosity of the region to be $\sim 3\times 10^{49}$ s$^{-1}$, using \citet{rubin68}. This corresponds to a single O5.5 V star, according to \citet{vaccaetal96}.  

We note that the XUV disk feature seen in NGC 4157 is similar to the Type 1 XUV disk structures as defined in \citet{thilkeretal07}. A study by \citet{lemoniasetal11} of 561 local galaxies found the average frequency of XUV disks to be between $4\%-14\%$. NGC 4157 is the only galaxy in the 35-galaxy CHANG-ES sample to show an XUV disk structure, which is consistent with the lower limit found in that study. An attempt was made to detect the XUV feature in the CHANG-ES radio continuum observations of this galaxy. However, if there is a corresponding feature in the radio continuum, the CHANG-ES data are insufficiently sensitive to trace the extended XUV feature or the dim corresponding {\HII} region. We estimated the $2\sigma$ upper limit for the average EM of the XUV disk region of the H$\alpha$ image (assuming it completely fills the radio beam) and found it corresponds to a minimum surface brightness limit of $\sim 2$ $\mu$Jy to detect C-band radio continuum emission from the XUV disk. 

Deep 21-cm {\HI} observations of NGC 4157 are outlined in \citet{kennedy09}. As seen in Figure 3.9 of that work, the neutral gas emission seems to extend slightly asymmetrically toward the northeast, which is the side of the outer disk hosting the anomalous {\HII} region. It is possible this gas asymmetry traces a recent accretion event onto the disk on NGC 4157 which led to the formation of stars at the location of the anomalous {\HII} region. 

It is interesting that the XUV feature and {\HI} emission is coincident with a star forming region at extreme distance from the disk. We theorize that the XUV disk could originate from a recent accretion event from a companion onto the disk of NGC 4157, and the extreme outer disk star formation region arose from gravitational instability within gas from that event \citep{fergusonetal98}. Potential future optical spectroscopic observations of the gas content surrounding this region are currently in the planning phase, which would help disentangle its origins. 

\section{Conclusions}

We present H$\alpha$ imaging to complement the CHANG-ES radio continuum survey sample of galaxies. Integrated H$\alpha$ values for the CHANG-ES sample from this study and the literature were combined with \textit{WISE} $22$ micron data to estimate improved SFRs in each galaxy using the revised method outlined in \citet{vargasetal18}. These SFRs were found to be in good agreement with the $22$ micron only SFR estimates from \citet{wiegertetal15}, after correcting those values for $22$ micron extinction. A newly discovered correlation between radio continuum scale heights from \citet{krauseetal18} and SFR and SFR$_{\rm{SD}}$ was found. Both the thermal and non-thermal component to radio continuum emission arise from star formation processes, which gives rise to this correlation. 

The galaxies with the largest deviations between their revised (mixture of H$\alpha$ + mid-IR) and their mid-IR only SFRs are the galaxies with the largest fractional H$\alpha$ contribution to the mixture. This, combined with the newly discovered correlation with radio scale height, suggests that a mid-IR extinction correction should be applied to SFR calibrations when used in edge-on galaxies, due to attenuation by dust. We also note that previous estimates of SFR in edge-on galaxies using mid-IR only calibrations without a correction for extinction may be inaccurate. 

Previously, it was not possible to identify the origins of the relationship between radio scale height and radio diameter without also seeing correlation between SFR and radio scale height. The updated SFR values from this study also show newly discovered correlation with radio scale length and radio diameter. The existence of these newly discovered correlations imply that the relationship between radio scale height and radio diameter originates from star formation within the disk.

Lastly, we identify a region of active star formation at extreme distance from the center of NGC 4157. We find this region is spatially coincident with an XUV disk feature, as found in GALEX NUV imaging, and is potentially ionized by a single O5.5 V star. 

A future study is planned involving the detailed fitting and measurement of H$\alpha$ scale heights, corrected with an appropriate dust extinction model dependent upon vertical height above the galaxy mid-plane.  

New H$\alpha$ images from this work can be found at the CHANG-ES data release web site, https://www.queensu.ca/changes

\acknowledgements
We thank the anonymous referee for thoughtful and helpful commentary on this manuscript. We also thank Rainer Beck for valuable comments to this manuscript. This material is based upon work supported by the National Science Foundation Graduate Research Fellowship under Grant No. 127229 to CJV. This material is also based on work partially supported by the National Science Foundation under Grant No. T-0908126, AST-1615594 to RAMW, and AST 1616513 to RJR and RAMW.

\newpage
\bibliography{refs_thermal.bib}
\bibliographystyle{apj}

\appendix

\begin{sidewaysfigure*}
\centering
\includegraphics[scale=0.39]{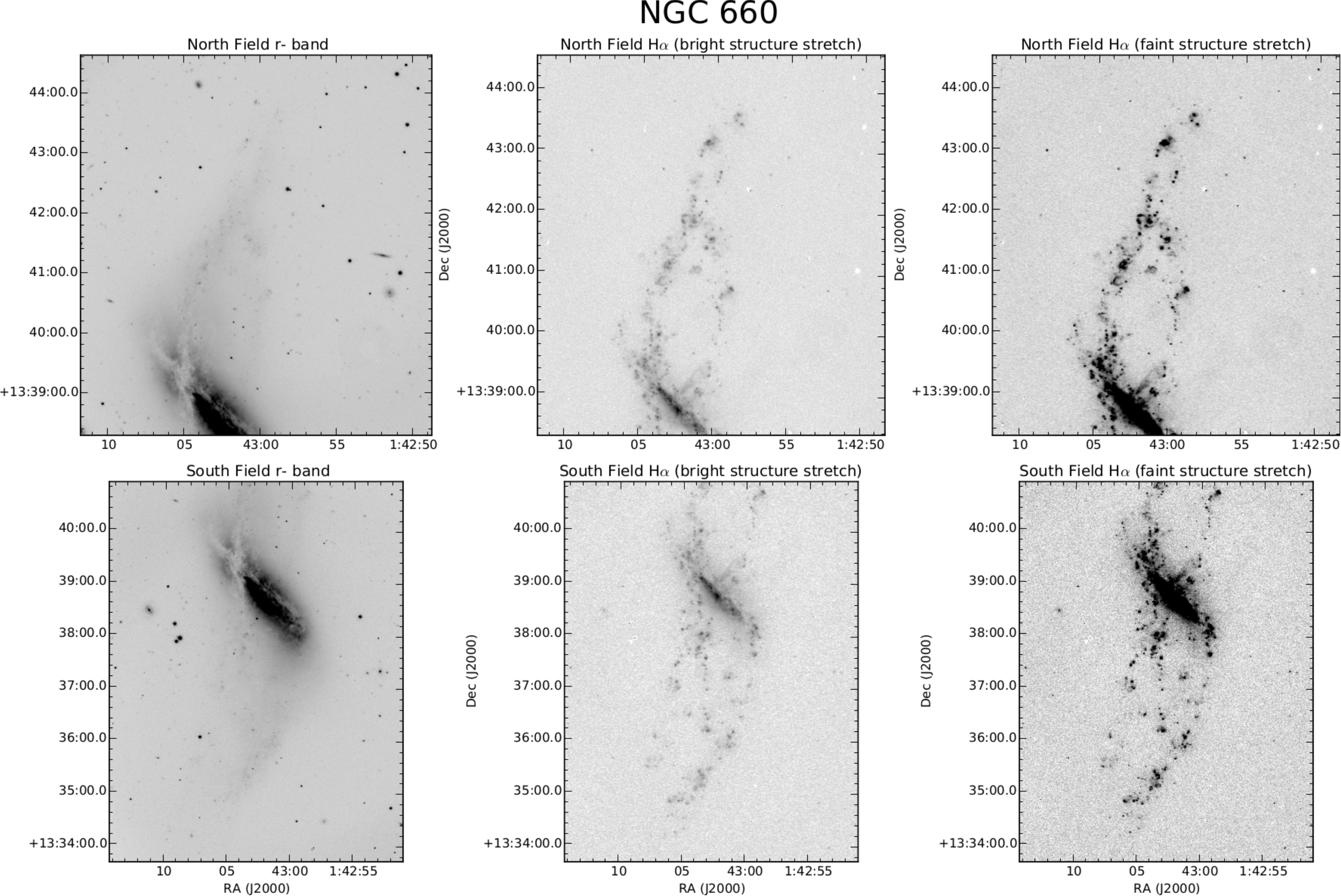}
\caption[NGC 660]{H$\alpha$ imaging for NGC 660. The upper row contains images of the northern field, and the lower row contains images of the southern field. The leftmost column contains the r-band continuum images. The central column contains the continuum subtracted H$\alpha$ images, shown in a stretch chosen to show bright disk features. The rightmost column contains the continuum subtracted H$\alpha$ images, shown in a stretch chosen to show faint H$\alpha$ structures. The minimum displayed pixel value in all H$\alpha$ images throughout this section corresponds to an EM of $-10$ pc cm$^{-6}$ so noise characteristics can be seen. The maximum displayed pixel value in the bright structure stretched H$\alpha$ images and faint structure stretched images correspond to an EM of $\sim3460$ pc cm$^{-6}$ and $\sim60$ pc cm$^{-6}$, respectively. All images throughout this section are shown in a logarithmic stretch.  }
\label{Haplot_660}
\end{sidewaysfigure*}

\begin{sidewaysfigure*}
\centering
\includegraphics[scale=0.39]{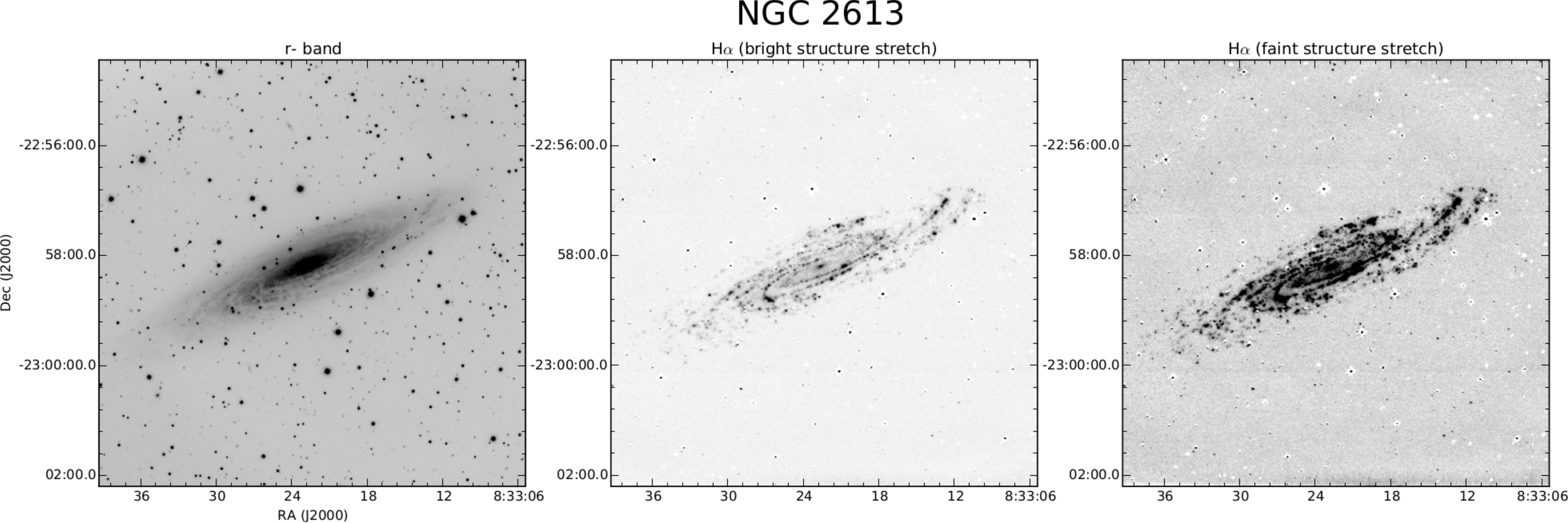}
\caption[NGC 2613]{H$\alpha$ imaging for NGC 2613. Left: r-band continuum image. Middle: continuum subtracted H$\alpha$ image, shown in a stretch chosen to show bright disk features. Left: continuum subtracted H$\alpha$ image, shown in a stretch chosen to show faint structures. All images are shown in a logarithmic stretch. The maximum displayed pixel value in the bright structure stretched H$\alpha$ images and faint structure stretched image correspond to an EM of $\sim415$ pc cm$^{-6}$ and $\sim75$ pc cm$^{-6}$, respectively.}
\label{Haplot_2613}
\end{sidewaysfigure*}

\begin{sidewaysfigure*}
\centering
\includegraphics[scale=0.39]{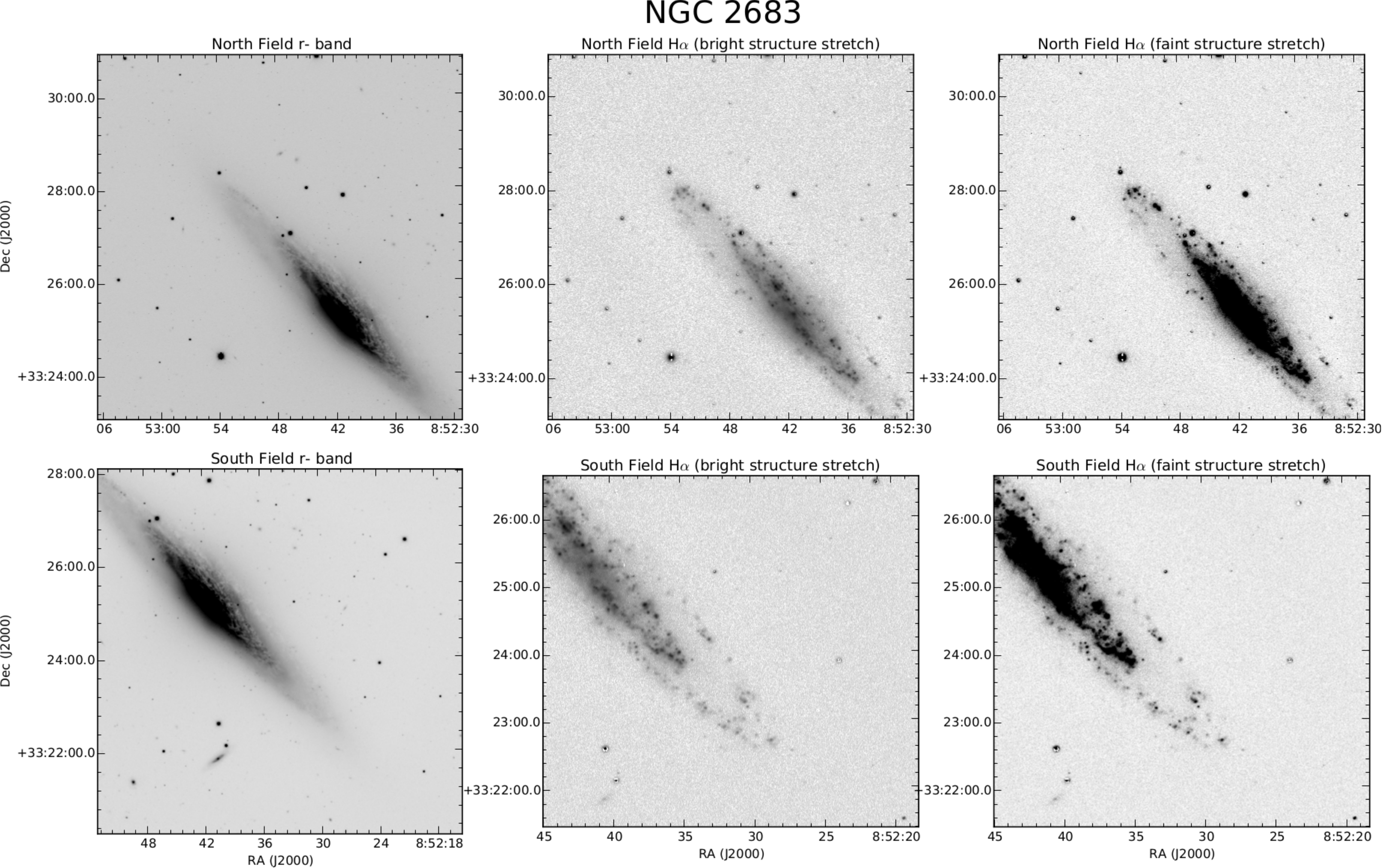}
\caption[NGC 2683]{H$\alpha$ imaging for NGC 2683. The maximum displayed pixel value in the bright structure stretched H$\alpha$ images and faint structure stretched images correspond to an EM of $\sim3825$ pc cm$^{-6}$ and $\sim200$ pc cm$^{-6}$, respectively. Other panel details are the same as in Figure \ref{Haplot_660}. }
\end{sidewaysfigure*}

\begin{sidewaysfigure*}
\centering
\includegraphics[scale=0.39]{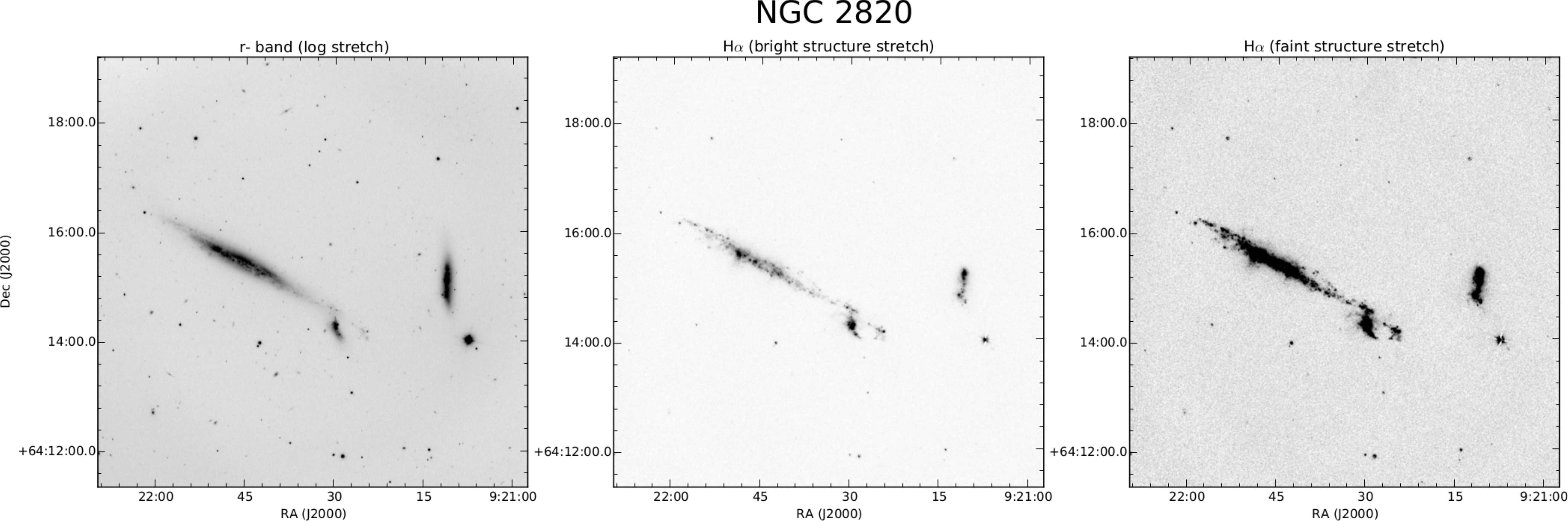}
\caption[NGC 2820]{H$\alpha$ imaging for NGC 2820. The maximum displayed pixel value in the bright structure stretched H$\alpha$ image and faint structure stretched image correspond to an EM of $\sim375$ pc cm$^{-6}$ and $\sim75$ pc cm$^{-6}$, respectively. Other panel details are the same as in Figure \ref{Haplot_2613} }
\label{Haplot_2820}
\end{sidewaysfigure*}

\begin{sidewaysfigure*}
\centering
\includegraphics[scale=0.39]{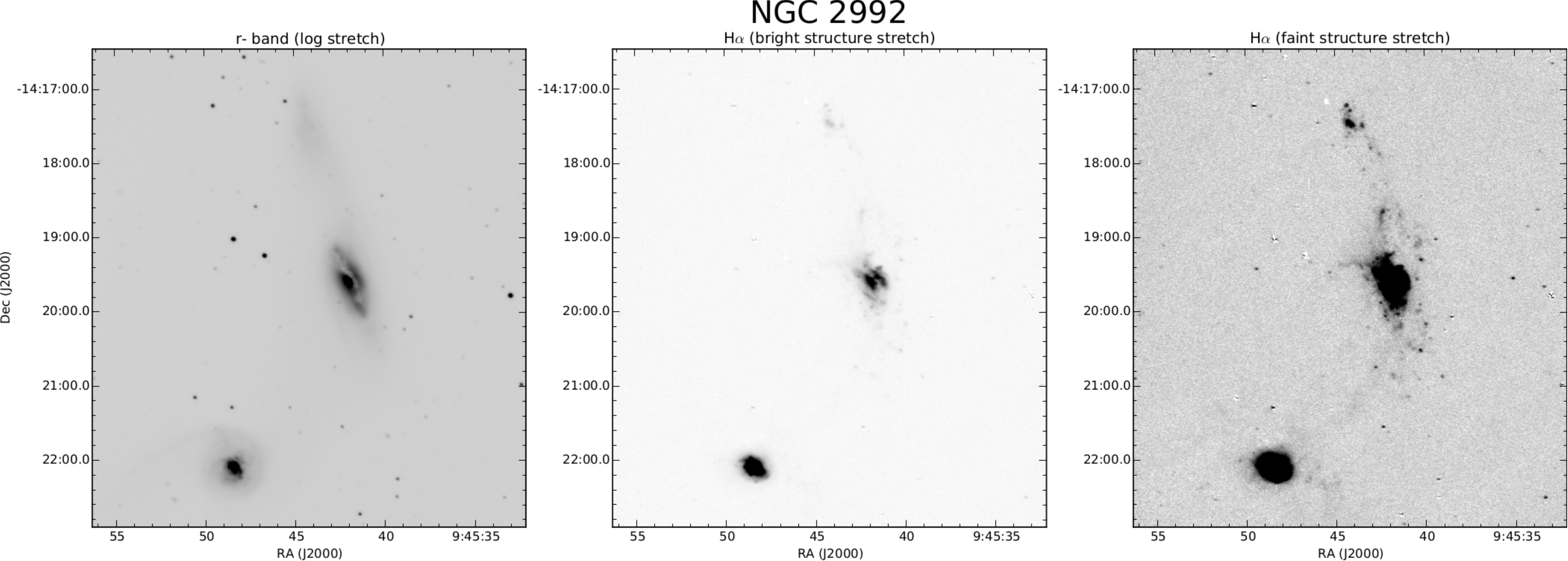}
\caption[NGC 2992]{H$\alpha$ imaging for NGC 2992. The maximum displayed pixel value in the bright structure stretched H$\alpha$ image and faint structure stretched image correspond to an EM of $\sim1345$ pc cm$^{-6}$ and $\sim62$ pc cm$^{-6}$, respectively. Other panel details are the same as in Figure \ref{Haplot_2613}.}
\label{Haplot_2992}
\end{sidewaysfigure*}

\begin{sidewaysfigure*}
\centering
\includegraphics[scale=0.39]{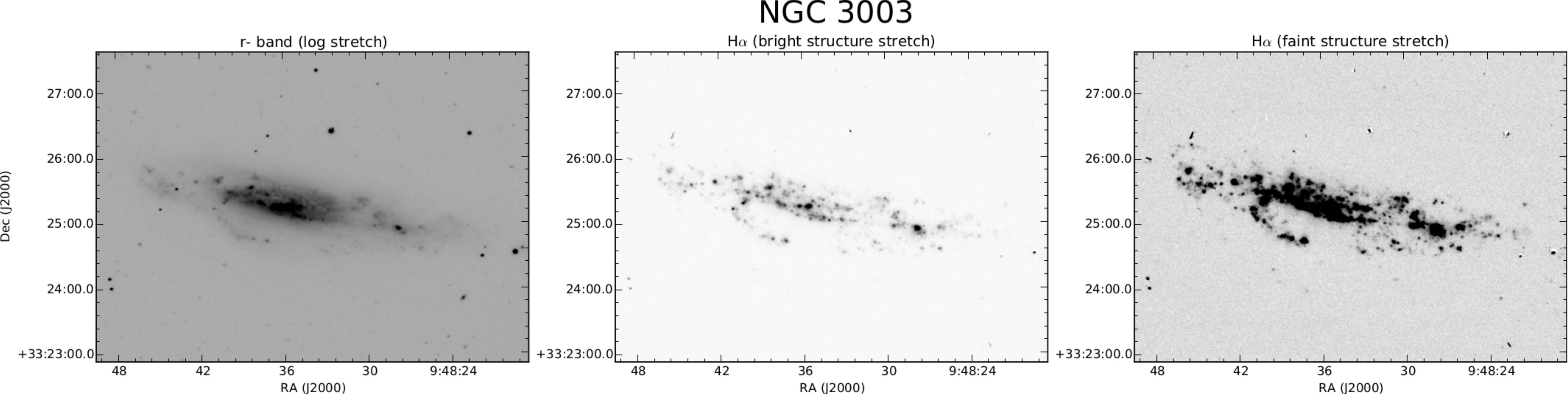}
\caption[NGC 3003]{H$\alpha$ imaging for NGC 3003. The maximum displayed pixel value in the bright structure stretched H$\alpha$ image and faint structure stretched image correspond to an EM of $\sim1845$ pc cm$^{-6}$ and $\sim115$ pc cm$^{-6}$, respectively. Other panel details are the same as in Figure \ref{Haplot_2613}.}
\label{Haplot_3003}
\end{sidewaysfigure*}

\begin{sidewaysfigure*}
\centering
\includegraphics[scale=0.39]{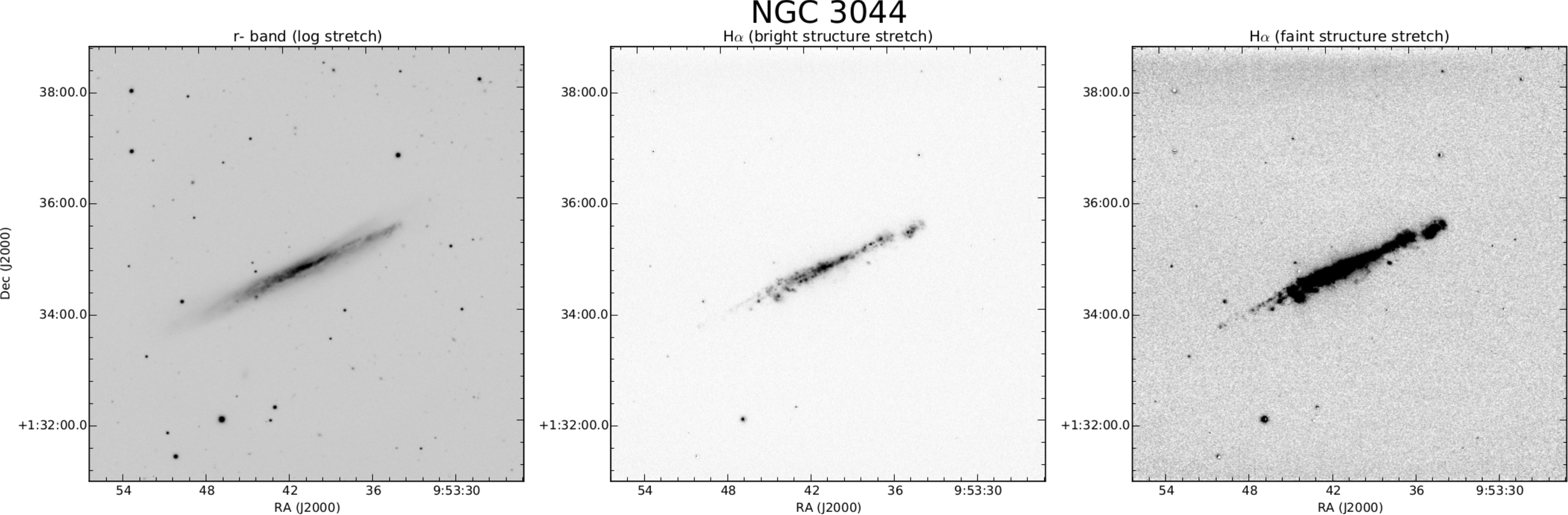}
\caption[NGC 3044]{H$\alpha$ imaging for NGC 3044. The maximum displayed pixel value in the bright structure stretched H$\alpha$ image and faint structure stretched image correspond to an EM of $\sim2035$ pc cm$^{-6}$ and $\sim120$ pc cm$^{-6}$, respectively. Other panel details are the same as in Figure \ref{Haplot_2613}.}
\label{Haplot_3044}
\end{sidewaysfigure*}

\begin{sidewaysfigure*}
\centering
\includegraphics[scale=0.39]{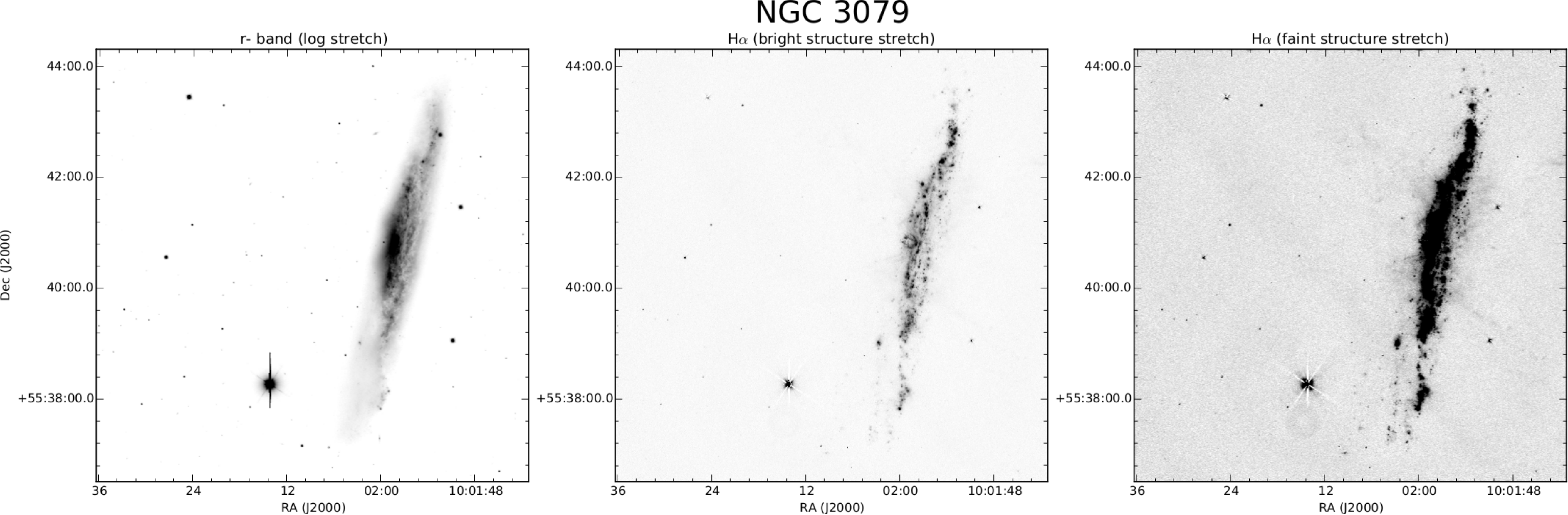}
\caption[NGC 3079]{H$\alpha$ imaging for NGC 3079. The maximum displayed pixel value in the bright structure stretched H$\alpha$ image and faint structure stretched image correspond to an EM of $\sim1555$ pc cm$^{-6}$ and $\sim115$ pc cm$^{-6}$, respectively. Other panel details are the same as in Figure \ref{Haplot_2613}.}
\label{Haplot_3079}
\end{sidewaysfigure*}

\begin{sidewaysfigure*}
\centering
\includegraphics[scale=0.39]{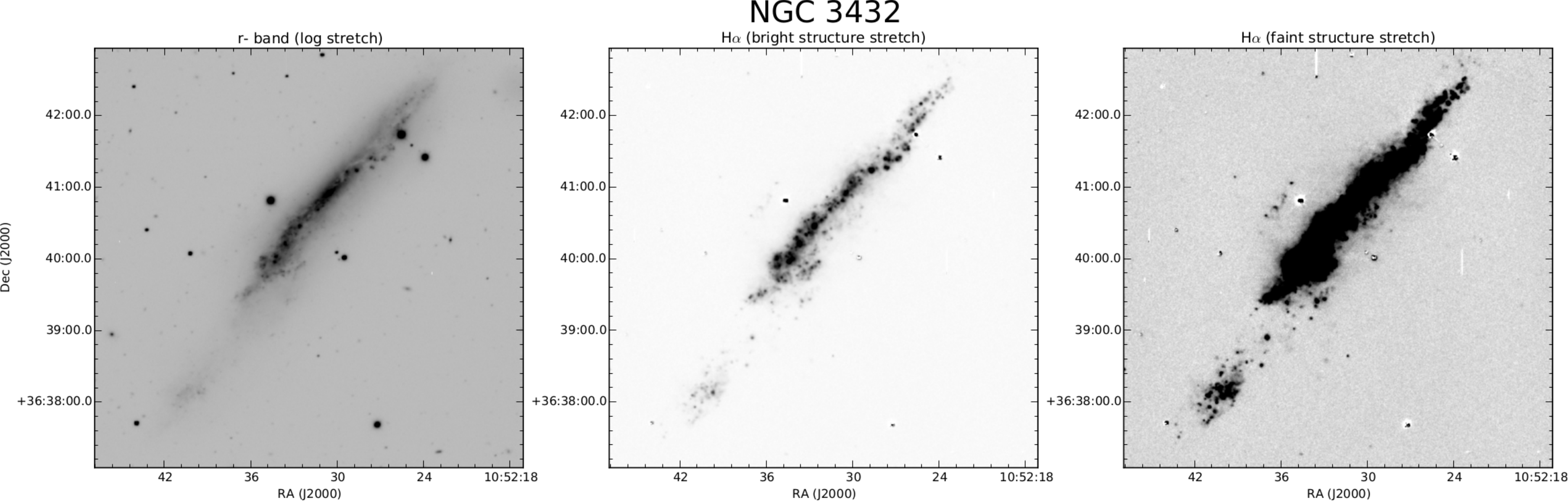}
\caption[NGC 3432]{H$\alpha$ imaging for NGC 3432. The maximum displayed pixel value in the bright structure stretched H$\alpha$ image and faint structure stretched image correspond to an EM of $\sim2170$ pc cm$^{-6}$ and $\sim75$ pc cm$^{-6}$, respectively. Other panel details are the same as in Figure \ref{Haplot_2613}.}
\end{sidewaysfigure*}

\begin{sidewaysfigure*}
\centering
\includegraphics[scale=0.39]{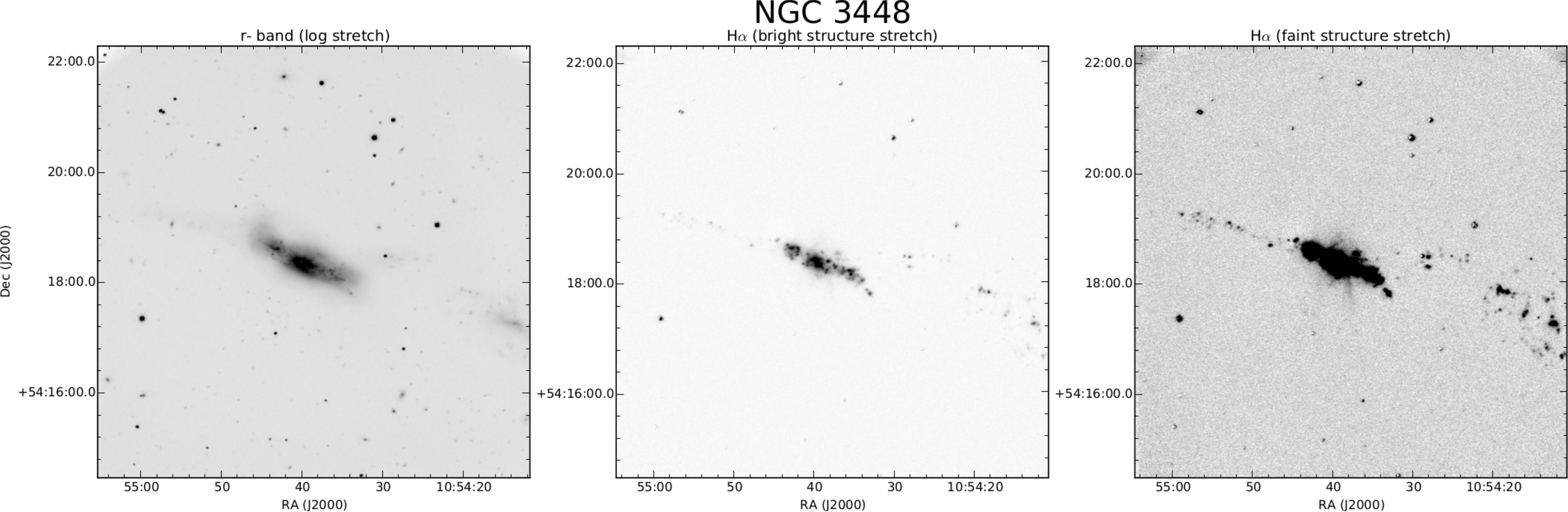}
\caption[NGC 3448]{H$\alpha$ imaging for NGC 3448. The maximum displayed pixel value in the bright structure stretched H$\alpha$ image and faint structure stretched image correspond to an EM of $\sim1540$ pc cm$^{-6}$ and $\sim70$ pc cm$^{-6}$, respectively. Other panel details are the same as in Figure \ref{Haplot_2613}.}
\end{sidewaysfigure*}

\begin{sidewaysfigure*}
\centering
\includegraphics[scale=0.39]{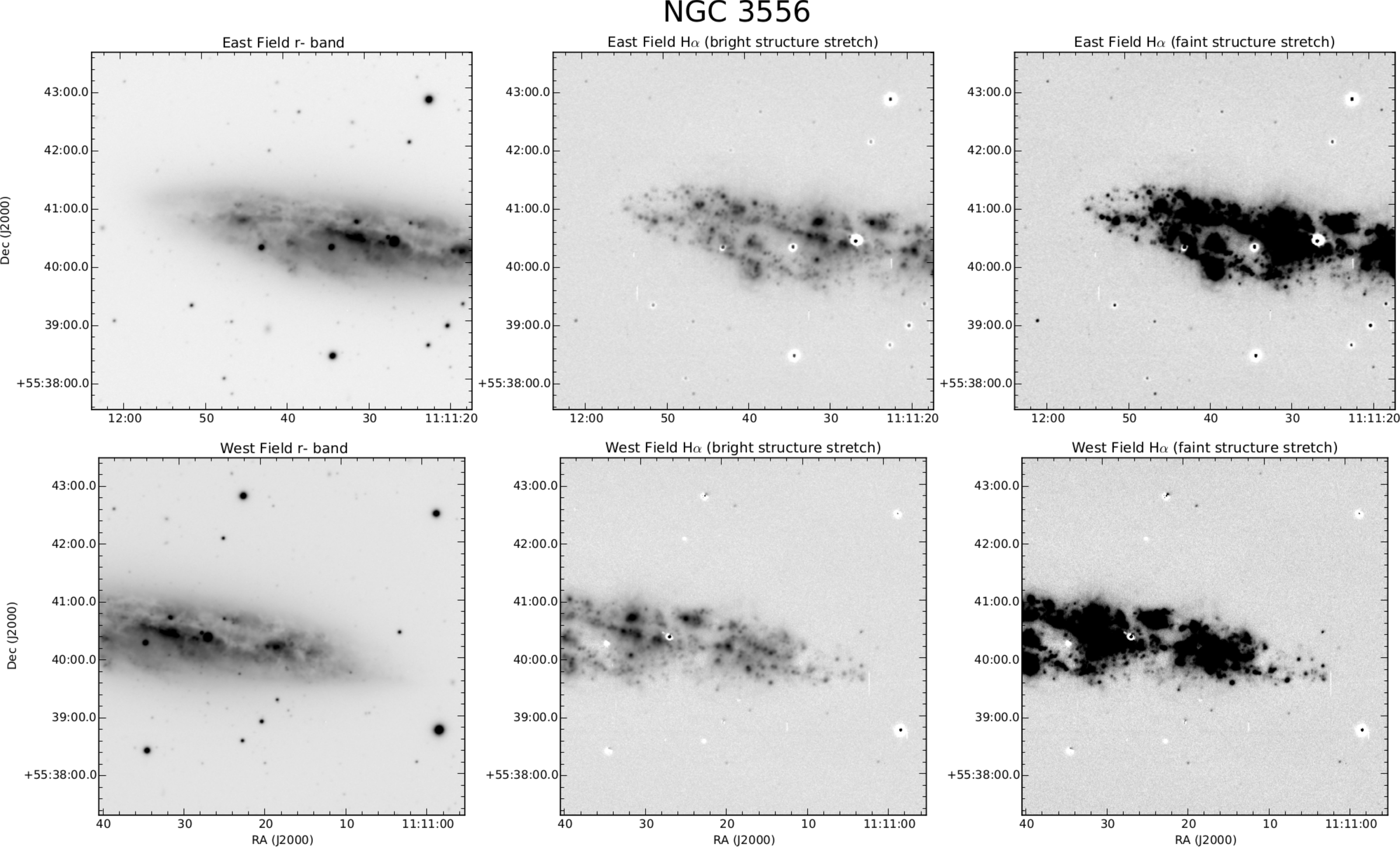}
\caption[NGC 3556]{H$\alpha$ imaging for NGC 3556. The maximum displayed pixel value in the bright structure stretched H$\alpha$ image and faint structure stretched image correspond to an EM of $\sim4030$ pc cm$^{-6}$ and $\sim$ pc cm$^{-6}$, respectively. Other panel details are the same as in Figure \ref{Haplot_660}.}
\end{sidewaysfigure*}

\begin{sidewaysfigure*}
\centering
\includegraphics[scale=0.44]{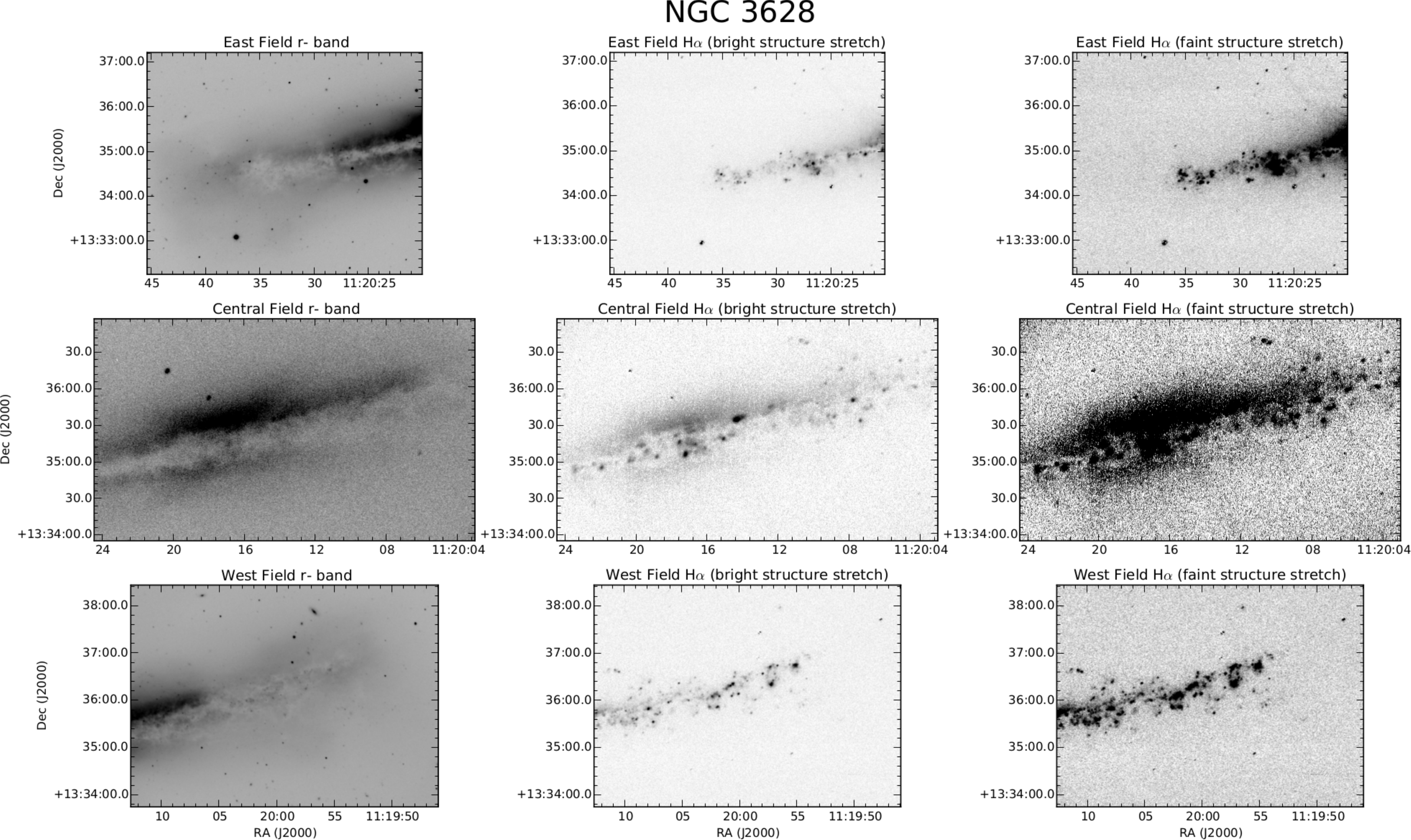}
\caption[NGC 3628]{H$\alpha$ imaging for NGC 3628. The maximum displayed pixel value in the bright structure stretched H$\alpha$ images and faint structure stretched images correspond to an EM of $\sim500$ pc cm$^{-6}$ and $\sim90$ pc cm$^{-6}$, respectively.}
\end{sidewaysfigure*}

\begin{sidewaysfigure*}
\centering
\includegraphics[scale=0.39]{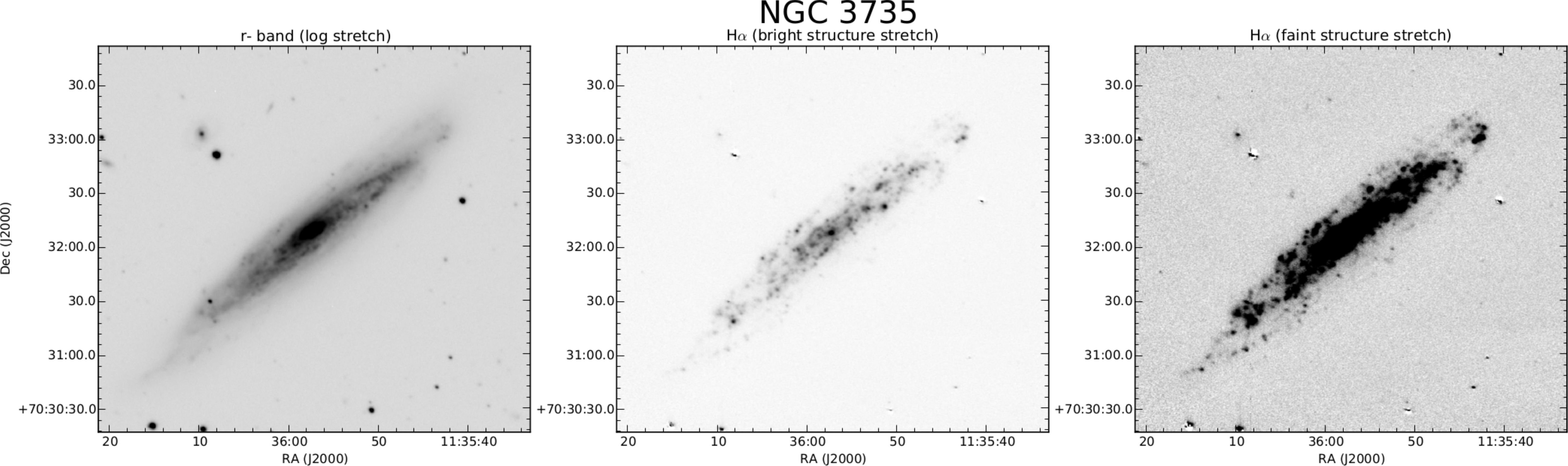}
\caption[NGC 3735]{H$\alpha$ imaging for NGC 3735. The maximum displayed pixel value in the bright structure stretched H$\alpha$ image and faint structure stretched image correspond to an EM of $\sim1070$ pc cm$^{-6}$ and $\sim75$ pc cm$^{-6}$, respectively. Other panel details are the same as in Figure \ref{Haplot_2613}.}
\end{sidewaysfigure*}

\begin{sidewaysfigure*}
\centering
\includegraphics[scale=0.39]{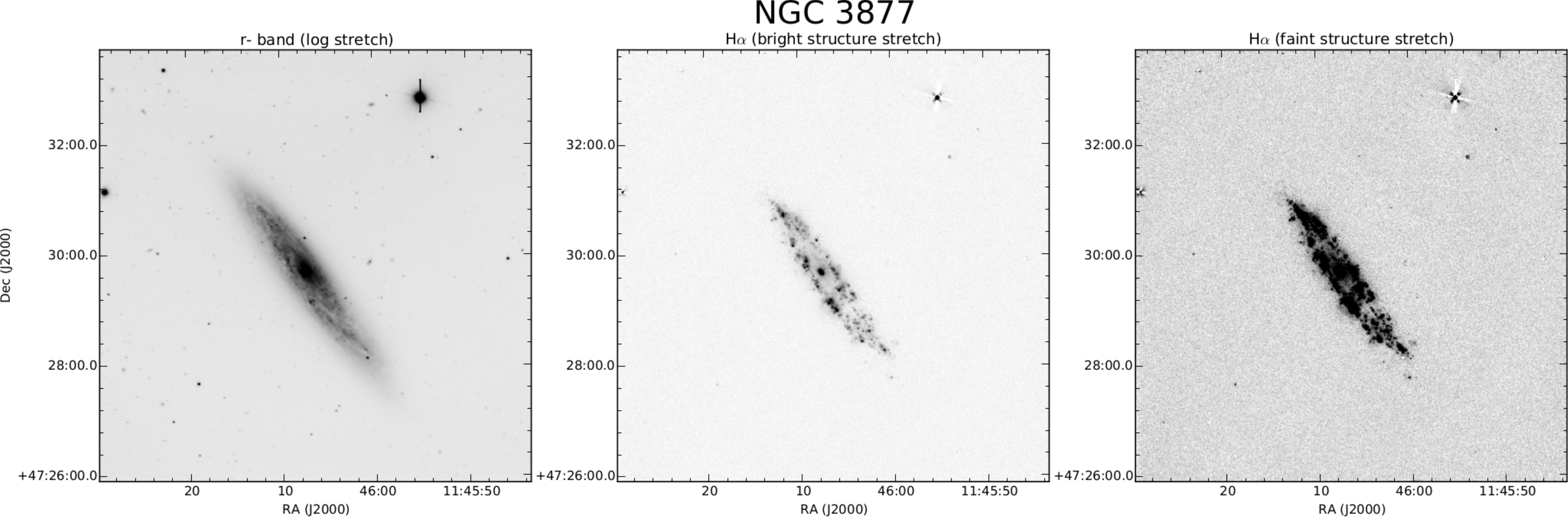}
\caption[NGC 3877]{H$\alpha$ imaging for NGC 3877. The maximum displayed pixel value in the bright structure stretched H$\alpha$ image and faint structure stretched image correspond to an EM of $\sim700$ pc cm$^{-6}$ and $\sim90$ pc cm$^{-6}$, respectively. Other panel details are the same as in Figure \ref{Haplot_2613}.}
\end{sidewaysfigure*}

\begin{sidewaysfigure*}
\centering
\includegraphics[scale=0.39]{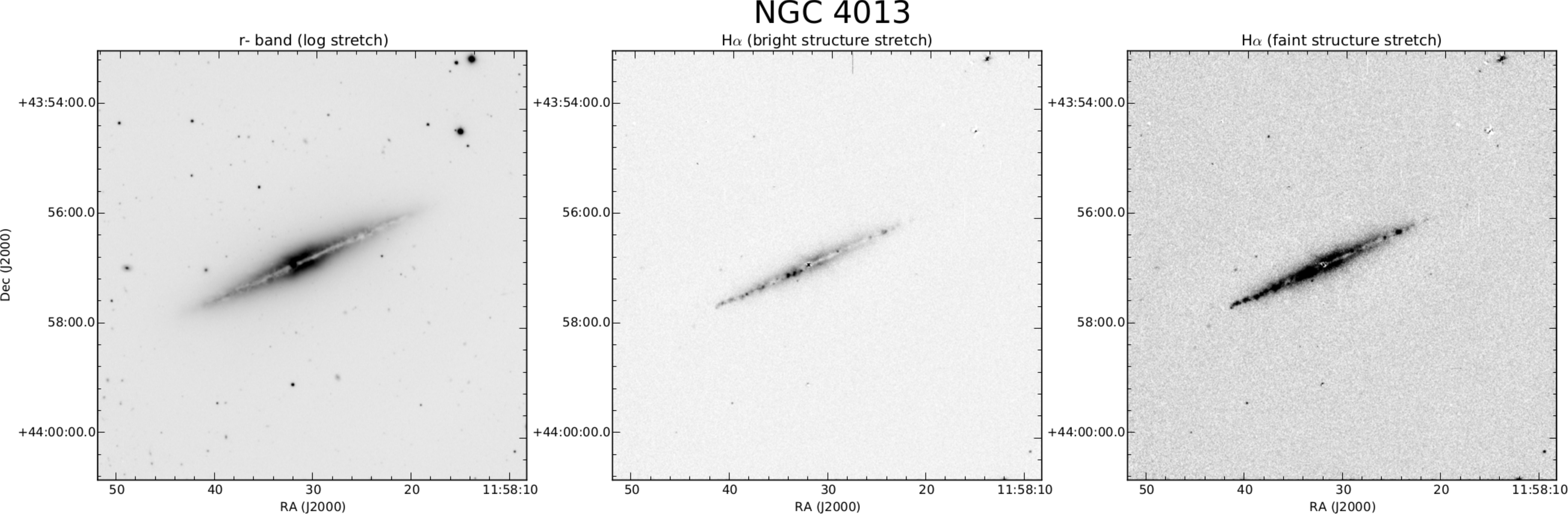}
\caption[NGC 4013]{H$\alpha$ imaging for NGC 4013. The maximum displayed pixel value in the bright structure stretched H$\alpha$ image and faint structure stretched image correspond to an EM of $\sim465$ pc cm$^{-6}$ and $\sim95$ pc cm$^{-6}$, respectively. Other panel details are the same as in Figure \ref{Haplot_2613}.}
\end{sidewaysfigure*}

\begin{sidewaysfigure*}
\centering
\includegraphics[scale=0.39]{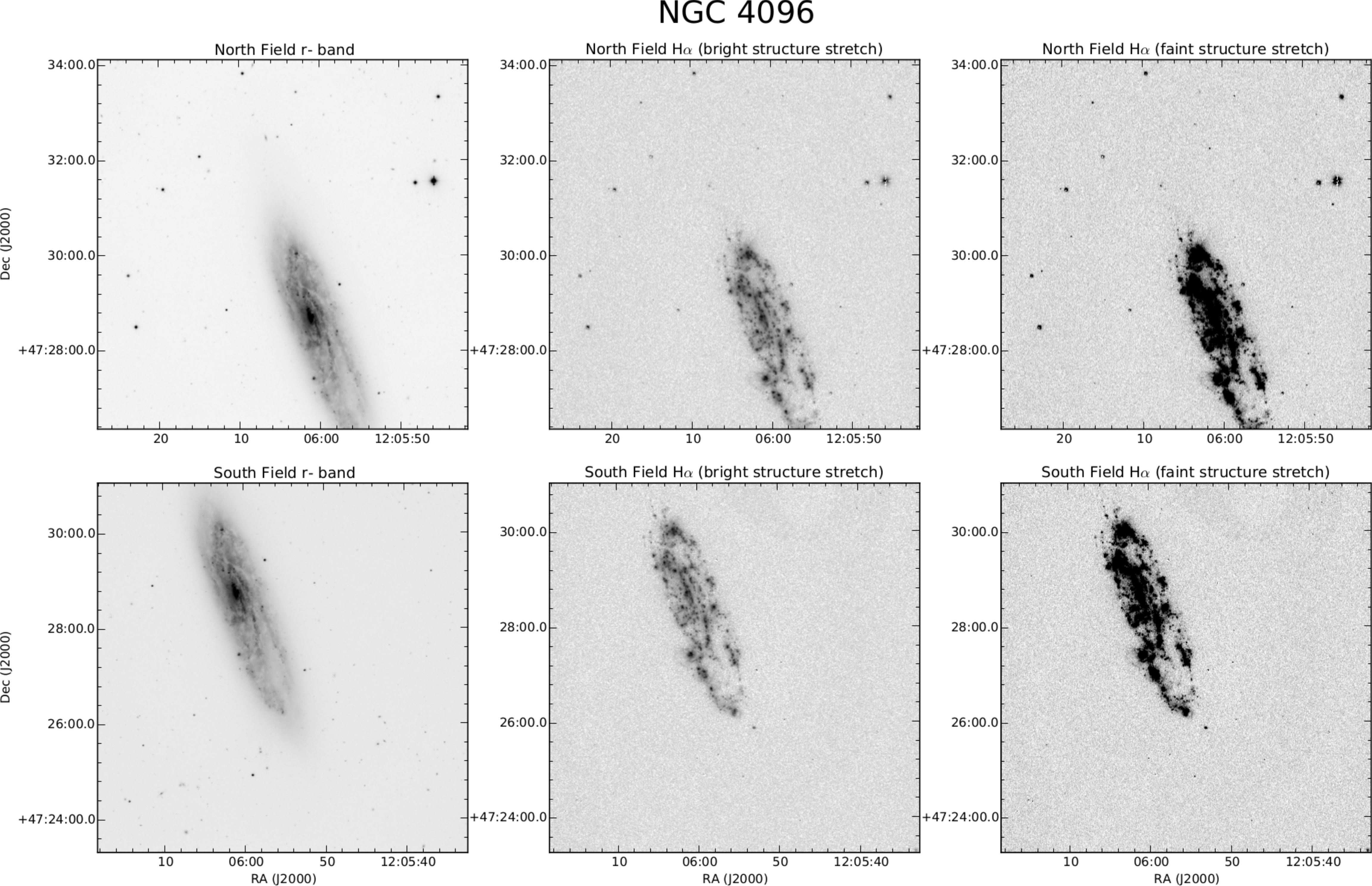}
\caption[NGC 4096]{H$\alpha$ imaging for NGC 4096. The maximum displayed pixel value in the bright structure stretched H$\alpha$ images and faint structure stretched images correspond to an EM of $\sim430$ pc cm$^{-6}$ and $\sim170$ pc cm$^{-6}$, respectively. Other panel details are the same as in Figure \ref{Haplot_660}.}
\end{sidewaysfigure*}

\begin{sidewaysfigure*}
\centering
\includegraphics[scale=0.39]{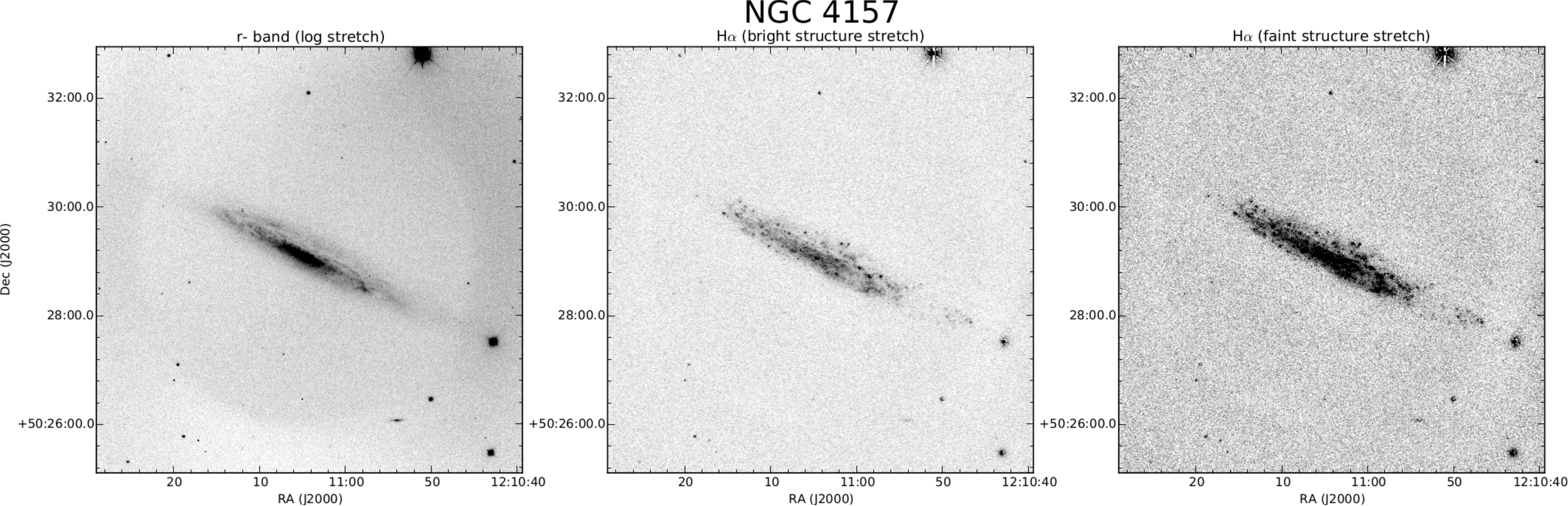}
\caption[NGC 4157]{H$\alpha$ imaging for NGC 4157. The maximum displayed pixel value in the bright structure stretched H$\alpha$ image and faint structure stretched image correspond to an EM of $\sim570$ pc cm$^{-6}$ and $\sim200$ pc cm$^{-6}$, respectively. Other panel details are the same as in Figure \ref{Haplot_2613}.}
\end{sidewaysfigure*}

\begin{sidewaysfigure*}
\centering
\includegraphics[scale=0.39]{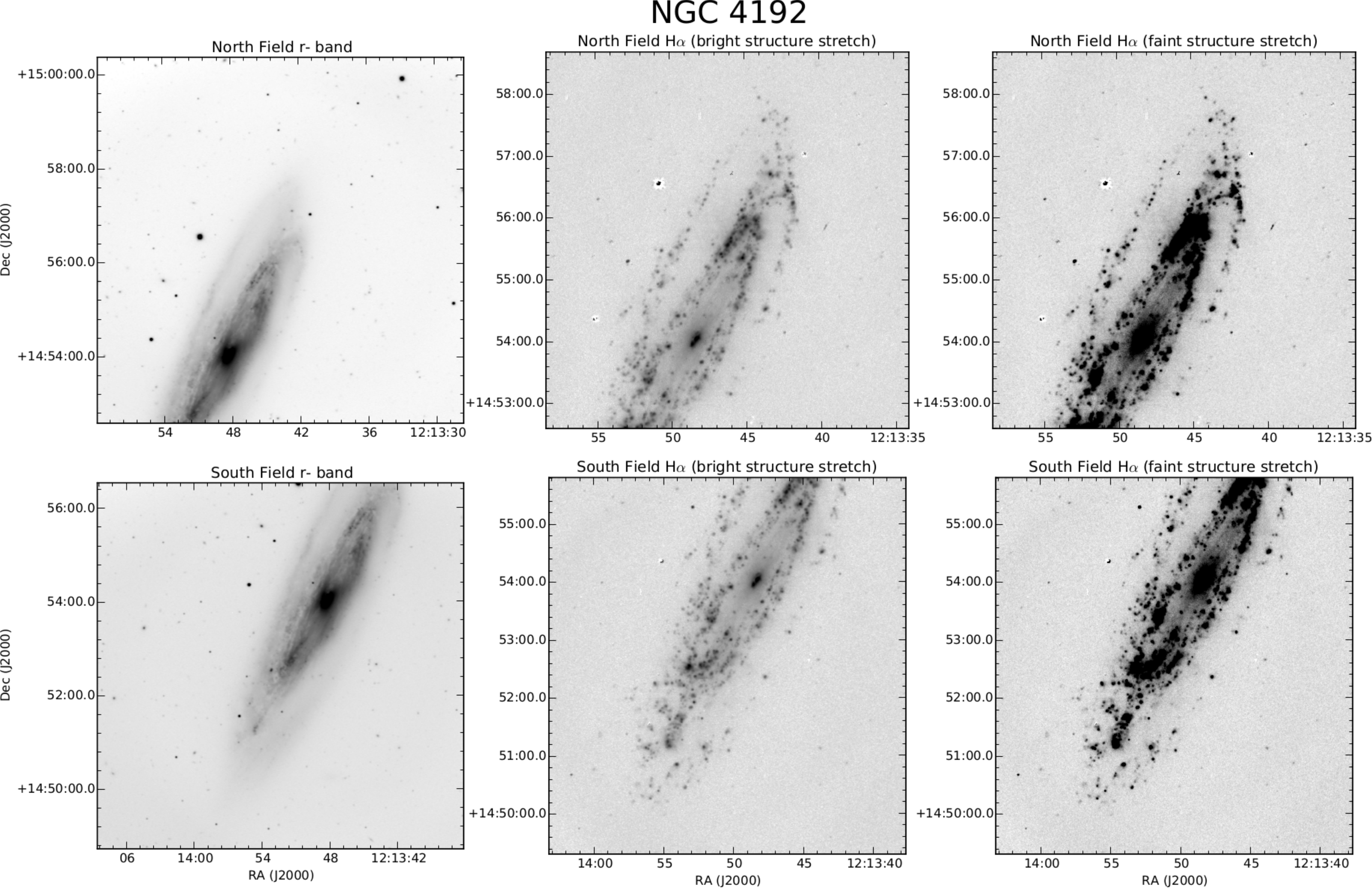}
\caption[NGC 4192]{H$\alpha$ imaging for NGC 4192. The maximum displayed pixel value in the bright structure stretched H$\alpha$ images and faint structure stretched images correspond to an EM of $\sim6340$ pc cm$^{-6}$ and $\sim260$ pc cm$^{-6}$, respectively. Other panel details are the same as in Figure \ref{Haplot_660}.}
\end{sidewaysfigure*}

\begin{sidewaysfigure*}
\centering
\includegraphics[scale=0.39]{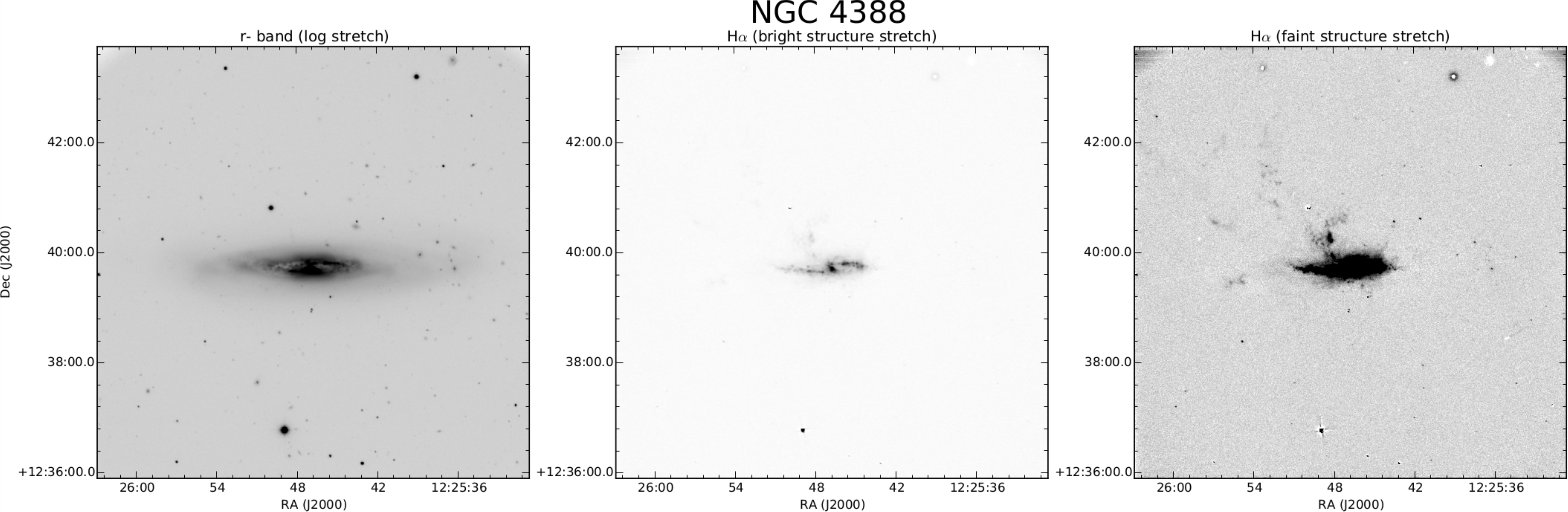}
\caption[NGC 4388]{H$\alpha$ imaging for NGC 4388. The maximum displayed pixel value in the bright structure stretched H$\alpha$ image and faint structure stretched image correspond to an EM of $\sim4000$ pc cm$^{-6}$ and $\sim70$ pc cm$^{-6}$, respectively. Other panel details are the same as in Figure \ref{Haplot_2613}.}
\end{sidewaysfigure*}

\begin{sidewaysfigure*}
\centering
\includegraphics[scale=0.41]{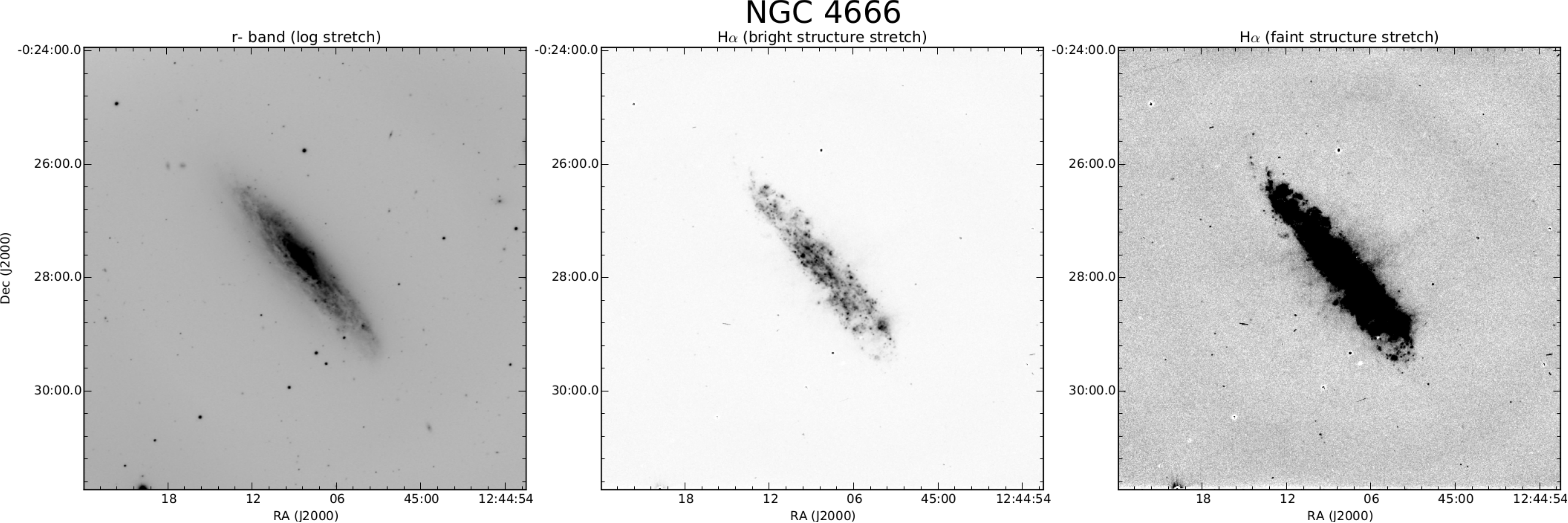}
\caption[NGC 4666]{H$\alpha$ imaging for NGC 4666. The maximum displayed pixel value in the bright structure stretched H$\alpha$ image and faint structure stretched image correspond to an EM of $\sim1710$ pc cm$^{-6}$ and $\sim50$ pc cm$^{-6}$, respectively. Other panel details are the same as in Figure \ref{Haplot_2613}.}
\end{sidewaysfigure*}

\begin{sidewaysfigure*}
\centering
\includegraphics[scale=0.39]{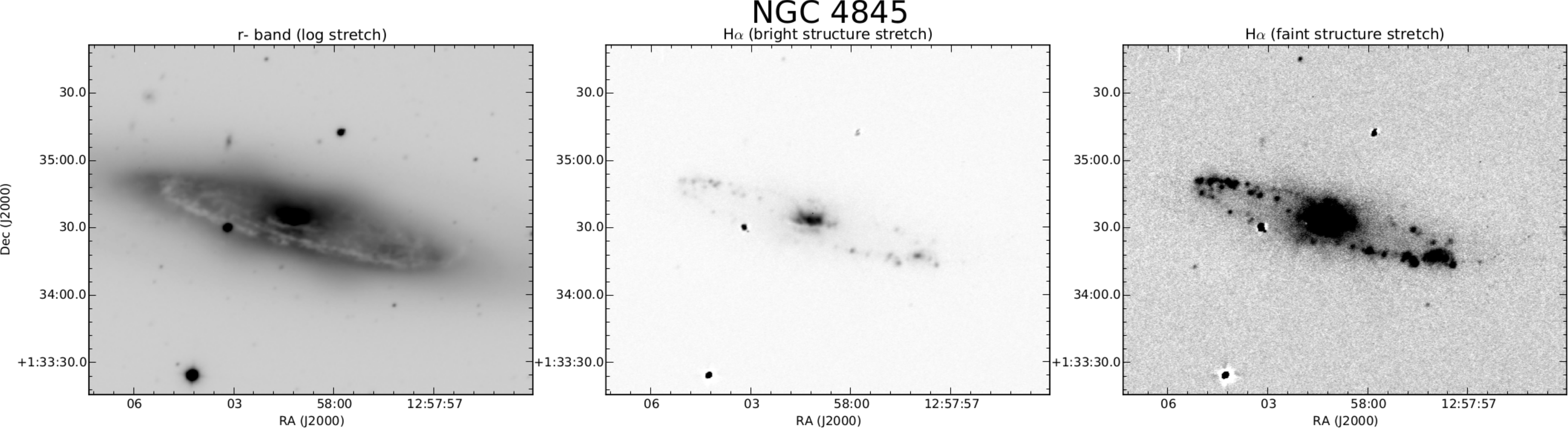}
\caption[NGC 4845]{H$\alpha$ imaging for NGC 4845. The maximum displayed pixel value in the bright structure stretched H$\alpha$ image and faint structure stretched image correspond to an EM of $\sim2085$ pc cm$^{-6}$ and $\sim65$ pc cm$^{-6}$, respectively. Other panel details are the same as in Figure \ref{Haplot_2613}.}
\end{sidewaysfigure*}

\begin{sidewaysfigure*}
\centering
\includegraphics[scale=0.39]{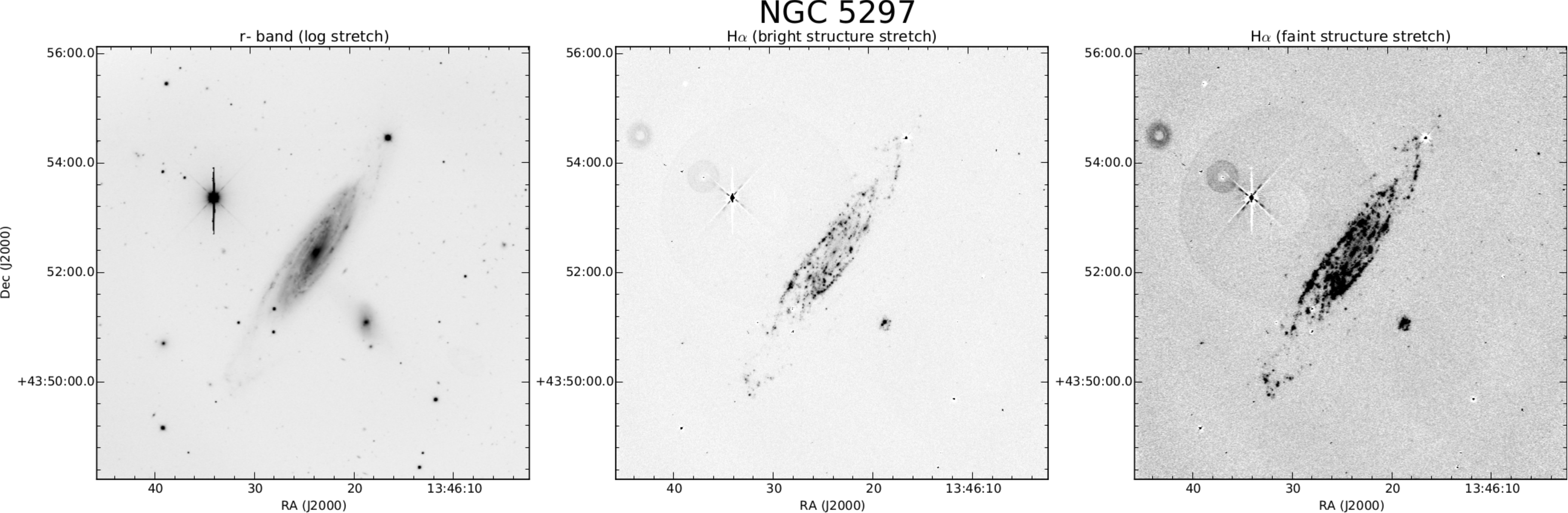}
\caption[NGC 5297]{H$\alpha$ imaging for NGC 5297. The maximum displayed pixel value in the bright structure stretched H$\alpha$ image and faint structure stretched image correspond to an EM of $\sim460$ pc cm$^{-6}$ and $\sim70$ pc cm$^{-6}$, respectively. Other panel details are the same as in Figure \ref{Haplot_2613}.}
\end{sidewaysfigure*}

\begin{sidewaysfigure*}
\centering
\includegraphics[scale=0.39]{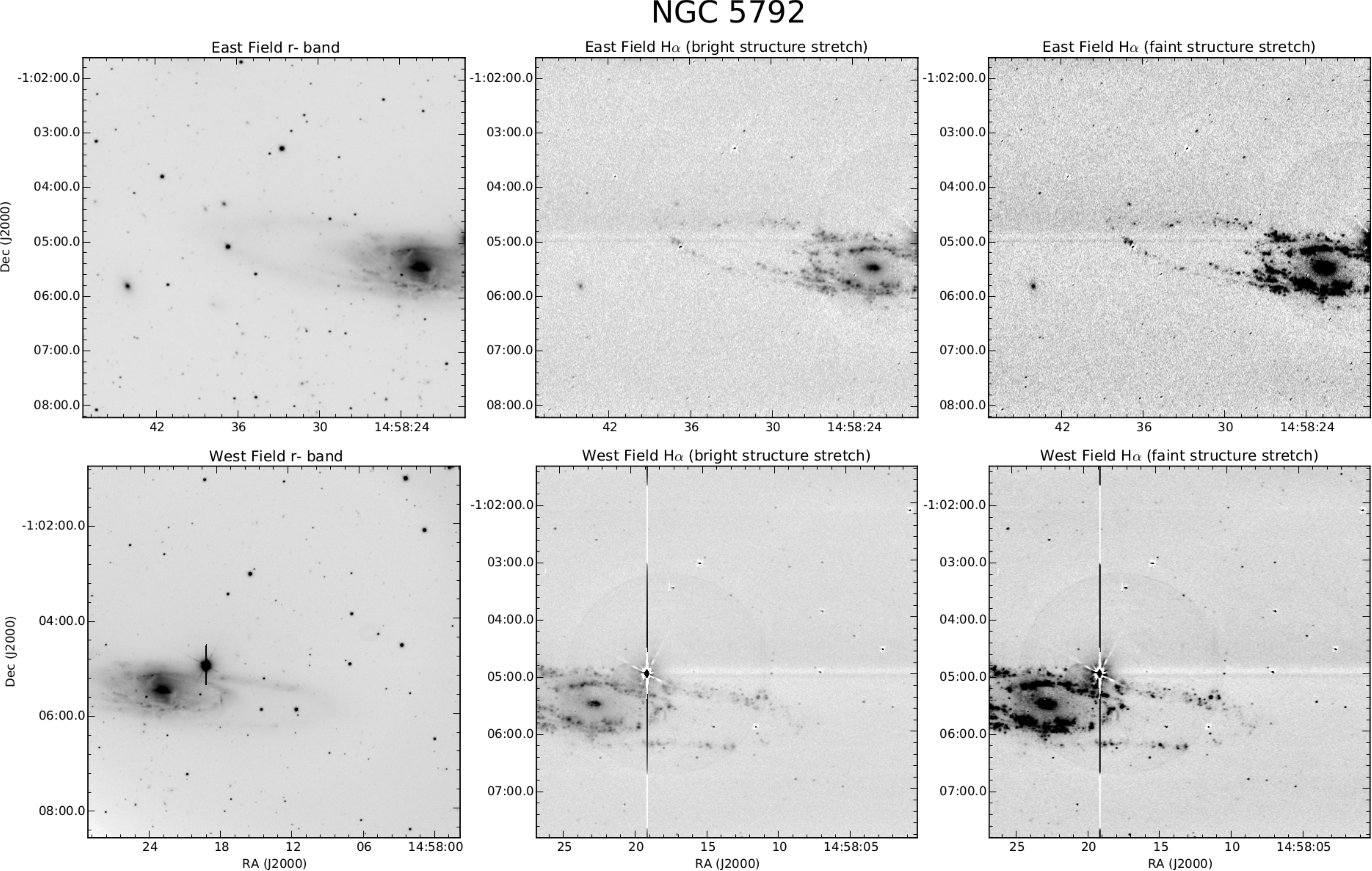}
\caption[NGC 5792]{H$\alpha$ imaging for NGC 5792. The maximum displayed pixel value in the bright structure stretched H$\alpha$ images and faint structure stretched images correspond to an EM of $\sim3080$ pc cm$^{-6}$ and $\sim110$ pc cm$^{-6}$, respectively. Other panel details are the same as in Figure \ref{Haplot_660}.}
\end{sidewaysfigure*}

\begin{sidewaysfigure*}
\centering
\includegraphics[scale=0.39]{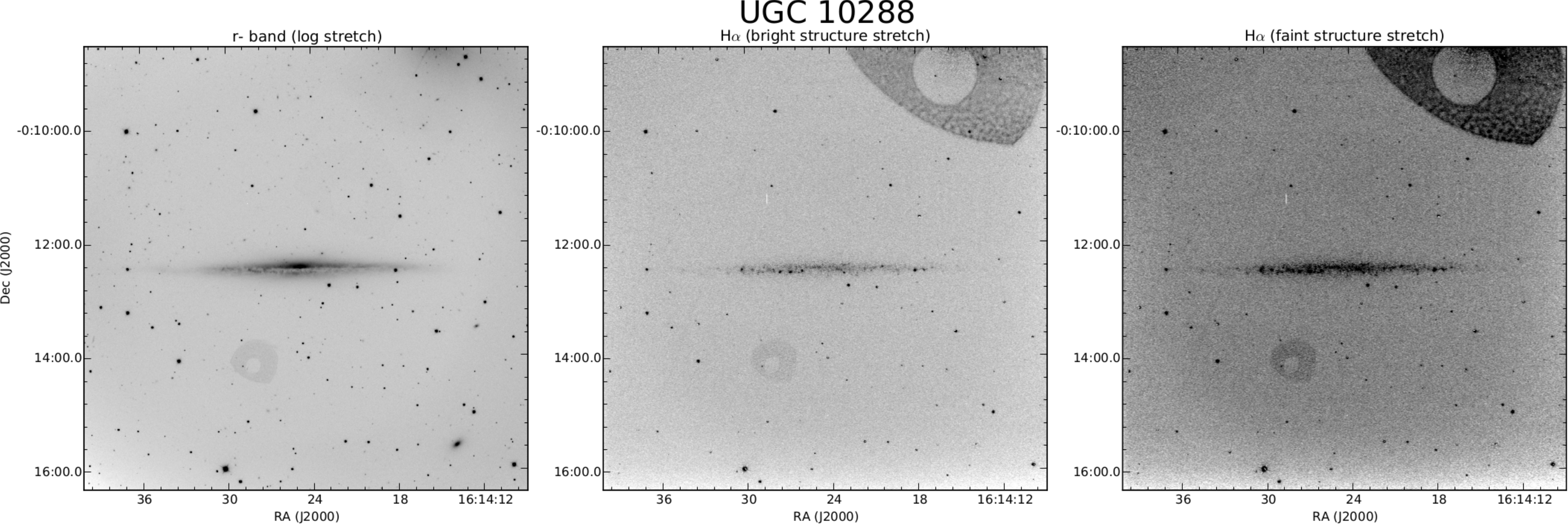}
\caption[UGC 10288]{H$\alpha$ imaging for UGC 10288. The maximum displayed pixel value in the bright structure stretched H$\alpha$ image and faint structure stretched image correspond to an EM of $\sim195$ pc cm$^{-6}$ and $\sim85$ pc cm$^{-6}$, respectively. Other panel details are the same as in Figure \ref{Haplot_2613}.}
\label{Haplot_10288}
\end{sidewaysfigure*}

\end{document}